\documentclass[10pt,final,journal]{IEEEtran}

\usepackage{amsmath,amssymb,amsfonts}
\usepackage{graphicx}
\usepackage{dblfloatfix}
\usepackage{siunitx}
\usepackage{support-caption}
\usepackage{subcaption}
\usepackage{tikz}
\usepackage{pgfplots}
\usepackage{pgfplotstable}
\pgfplotsset{compat=1.7}
\usepackage[normalem]{ulem}
\usepgfplotslibrary{
    groupplots,
}
\pgfplotsset{/pgfplots/group/.cd,
    horizontal sep=0.25cm,
    vertical sep=0.5cm
}
\usepackage[skins]{tcolorbox}
\pgfplotsset{
    compat=1.13,
    my style/.style={
        width=4.5cm,height=4.5cm,
        xlabel={Bitrate (bpp)},
        ylabel={Distortion (PSNR)},
        xmin=0,xmax=5
    },
    my style1/.style={
        width=4.5cm,height=4.5cm,
        xlabel={Bitrate (bpp)},
        ylabel={Max. abs. error ($\ell_\infty$)},
        xmin=0,xmax=5
    },
    residual legend style/.style={
        legend entries={
            $\ell_\infty^{4d}$DNet,
            EPIC \cite{mukati2020epipolar},
            $\ell_\infty$-SDNet \cite{zhang2020ultra},
            $\ell_\infty$-SASNet \cite{jin2020learning},
            HEVC,
            MuLE \cite{schelkens2019jpeg}
        },
        legend style={
            legend cell align={left},
            at={([yshift=0pt]0,1.15)},
            anchor=south west,
            /tikz/every even column/.append style={column sep=0.6cm}
        },
        legend columns=4,
    },
    residual legend style1/.style={
        legend entries={
            $\ell_\infty^{4d}$DNet,
            EPIC \cite{mukati2020epipolar},
            HEVC,
            MuLE \cite{schelkens2019jpeg}
        },
        legend style={
            legend cell align={left},
            at={([yshift=0pt]0,1.15)},
            anchor=south west,
            /tikz/every even column/.append style={column sep=1.3cm}
        },
        legend columns=4,
    },
}
\usetikzlibrary{spy,calc}
\usepackage{tabularx, booktabs}
\usepackage{diagbox}

\graphicspath{{images/}}

\definecolor{clr_hevc}{rgb}{0.0,0.0,0.0}
\definecolor{clr_x265}{rgb}{0.9922,0.7529,0.5255}
\definecolor{clr_plt1}{rgb}{0.894,0.101,0.109}
\definecolor{clr_plt2}{rgb}{0.215,0.494,0.721}
\definecolor{clr_plt3}{rgb}{0.301,0.686,0.290}
\definecolor{clr_plt4}{rgb}{0.651,0.337,0.157}
\definecolor{clr_plt5}{rgb}{0.596,0.305,0.639}
\definecolor{clr_plt6}{rgb}{1.0000,0.4980,0.000}

\newcommand{\norm}[1]{\left\lVert#1\right\rVert}

\begin{document}

\title{Deep Decoding of $\ell_\infty$-coded Light Field Images}

% Authors' detail
\author{M.~Umair~Mukati,~Xi~Zhang,~Xiaolin~Wu,~\IEEEmembership{Fellow,~IEEE,} and~S{\o}ren~Forchhammer,~\IEEEmembership{Member,~IEEE}% <-this % stops a space
\thanks{M. U. Mukati and S. Forchhammer are with the Department of Photonics Engineering, Technical University of Denmark, 2800 Kgs. Lyngby, Denmark e-mail: mummu, sofo @fotonik.dtu.dk.}
\thanks{X. Zhang is with the Department of Electronic Engineering, Shanghai Jiao Tong University, Shanghai, China (email: zhangxi\_19930818@sjtu.edu.cn).}
\thanks{X. Wu is with the Department of Electrical $\&$ Computer Engineering, McMaster University, Hamilton, L8G 4K1, Ontario, Canada (email: xwu@ece.mcmaster.ca).}% <-this % stops a space
\thanks{This project has received funding from EU's H2020 ITN programme, under the MSCA grant agreement No 765911 (RealVision).}
}

% Must be changed
% \markboth{Journal of \LaTeX\ Class Files,~Vol.~14, No.~8, August~2015}%
\markboth{MANUSCRIPT SUBMITTED TO IEEE TRANSACTIONS ON IMAGE PROCESSING}%
{Mukati \MakeLowercase{\textit{et al.}}: Deep Decoding of $\ell_\infty$-coded Light Field Images}

% Make the title area
\maketitle

\begin{abstract}
To enrich the functionalities of traditional cameras, light field cameras record both the intensity and direction of light rays, so that images can be rendered with user-defined camera parameters via computations.
The added capability and flexibility are gained at the cost of gathering typically more than $100\times$ greater amount of information than conventional images.
To cope with this issue, several light field compression schemes have been introduced. However, their ways of exploiting correlations of multidimensional light field data are complex and are hence not suited for inexpensive light field cameras.
In this work, we propose a novel $\ell_\infty$-constrained light-field image compression system that has a very low-complexity DPCM encoder and a CNN-based deep decoder. Targeting high-fidelity reconstruction, the CNN decoder capitalizes on the $\ell_\infty$-constraint and light field properties to remove the compression artifacts and achieves
significantly better performance than existing state-of-the-art $\ell_2$-based light field compression methods.
\end{abstract}


\IEEEpeerreviewmaketitle

\section{Introduction}
\label{sec:introduction}
\IEEEPARstart{T}{he} term light field refers to a set of light rays. Capturing a light field instead of an image of a scene offers several post-capture possibilities,
such as refocusing, aperture adjustment, view-point shifting, etc. \cite{ng2005light}.
A light ray in a light field can be parameterized by its interaction with two parallel planes, i.e., camera plane and image plane.
Hence, it can be represented in a 4D format as $\mathcal{L}\left(s,t,u,v\right)$, i.e., its intersection with the camera plane as its angular coordinates $\left(s,t\right)$, while intersection with the image plane as spatial coordinates $\left(u,v\right)$. In practical design, these coordinates can be recorded only at a fixed number of locations. In the camera plane, the number of intersection points determines the angular resolution, whereas the intersection points on the image plane determine spatial resolution.

A consumer-graded light field camera, Lytro Illum \cite{rerabek2015iso}, is based on plenoptic camera design, that places a micro-lens array in front of a photosensor in a traditional camera setup to record a scene from different perspectives. This camera design shares the photosensor's capacity to accommodate spatial and angular resolutions. A rectified light field obtained using this camera consumes as much as ${\sim}175$MB of storage space. However, due to the thin-baseline between intermediate perspectives, strong correlations exist among different views. To exploit the inter-view statistical redundancies, numerous compression schemes (lossy or lossless) have been developed for plenoptic light field image compression.
Depending on the application, either lossless or lossy compression mode can be adopted. Most of the compression schemes are designed to operate on the rectified light field rather than raw lenslet captures from the plenoptic cameras \cite{conti2020dense}.

For lossy compression of light field images, the earlier works simply utilized the advanced video codecs such as HEVC. The light field is presented to the HEVC encoder as a pseudo video sequence \cite{vieira2015data,li2016compression,ahmad2017interpreting}.
Recently, the JPEG committee initiated JPEG-Pleno to standardize the advances in this field. They provide two coding tools as standard schemes for light field compression \cite{schelkens2019jpeg}, one based on 4D DCT (MuLE) \cite{de20184d} and the another on 4D prediction (WaSP) \cite{astola2018wasp}.
To fully exploit correlations in light field signals for compression gains, the mainstream lossy methods employ a heavy-duty encoder, such as the ones reported in \cite{conti2020dense}. To reduce the encoding complexity, a few distributed light field coding schemes at the cost of high decoder complexity have been recently proposed \cite{mukati2020view, phicong2020adaptive, mukati2021improved}.

Lossless compression of light fields is a
more challenging task as the design objective becomes the exact reconstruction while approaching the entropy bound.  Due to the 4D nature of light fields, completely removing statistical redundancies in complex structures incurs a very high computational cost at the encoder side \cite{helin2017minimum, schiopu2017lossless, santos2018lossless}.
Such a heavy-duty encoder cannot be afforded by inexpensive light field cameras and is not suited for real-time applications.
In addition to high encoder complexity, strictly lossless coding of light fields only offers quite limited savings of storage and bandwidth due to the entropy bound.

To overcome the above weaknesses, we propose a novel asymmetric light-field image compression system consisting of a low-complexity $\ell_\infty$-constrained DPCM encoder and a CNN-based deep decoder.
The $\ell_\infty$-constrained (a.k.a., near-lossless) coding is a strategy to break the barrier of the source entropy while maintaining high data fidelity \cite{wu2000sub}.
It allows the
reconstruction error but imposes a tight $\ell_\infty$ bound $\tau$, i.e., the value of each decompressed pixel does not differ from its original value by more than a user-specified bound $\tau$.
The CNN deep decoder capitalizes on the known error bound $\tau$ and the 4D light field structures to remove the compression artifacts; this decoder network, denoted by $\ell_\infty^{4d}$DNet, is the main contribution of this work.
The $\ell_\infty^{4d}$DNet combines the best of $\ell_\infty$ and $\ell_2$ distortion metrics and achieves
significantly better $\ell_2$ performance than existing state-of-the-art light field compression methods, while ensuring a reconstruction error bound of $2\tau$ for every sample.

The design principle of the proposed $\ell_\infty^{4d}$DNet based light field coding system is similar to that of distributed source coding.  Highly correlated signal components of light field are encoded separately instead of jointly; the unexploited statistical redundancies are removed by a more sophisticated decoder to achieve good rate-distortion performance.  The purpose of the $\ell_\infty^{4d}$DNet deep decoder is to solve the inverse problem of estimating the latent original light field from an $\ell_\infty$-compressed light field.
Although $\ell_\infty^{4d}$DNet is a heavy-duty decoder, it is coupled with a streamlined real-time encoder; hence the system is suited for low-cost light field cameras.

The remainder of this paper is organized as follows. In Section \ref{sec:relevant}, we review related work on deep decompression of near-lossless images and light field refinement techniques. In Section \ref{sec:overview}, after introducing necessary notations, we give an overview of the $\ell_\infty^{4d}$DNet based light field compression system.  Section \ref{sec:slfdnet} presents and explains the architecture of $\ell_\infty^{4d}$DNet and the deep decompression process. Section \ref{sec:analysis} analyzes the performance of the $\ell_\infty^{4d}$DNet deep decoder against the state-of-the-art light field compression schemes; also, the performance of $\ell_\infty^{4d}$DNet is compared with those of alternate network architectures for deep light field decompression. Finally, the paper is concluded in Section \ref{sec:conclusion}.

\section{Background and relevant work}
\label{sec:relevant}
\noindent
Although there are several methods for removing compression artifacts of images and videos \cite{zhang2020ultra}, almost none of them is optimized for light fields.
To the best of the author's knowledge, the problem of compression artifacts removal in the context of light fields, let alone from $\ell_\infty$-constrained compressed light field, has not been addressed explicitly.
In this section,
we give a brief review of literature on topics related to this work, through which we also provide  necessary background information for readers to better follow our technical development in subsequent sections.

\subsection{Restoration of near-lossless coded images}
\noindent
Traditional methods for $\ell_\infty$ near-lossless image compression can be classified into two categories:
\begin{enumerate}
    \item Pre-quantization: Quantizing pixel values according to the $\ell_\infty$ error bound, and then losslessly compressing the pre-quantized pixels, e.g., near-lossless WebP~\cite{webp};
    \item Predictive coding: predicting current pixels based on previously compressed pixels, then quantizing the prediction residuals to satisfy the $\ell_\infty$ error bound, and finally encoding the quantized prediction residuals, e.g., \cite{chen1994near,ke1998near}, near-lossless JPEG-LS \cite{weinberger2000loco} and near-lossless CALIC \cite{wu2000sub}.
\end{enumerate}

To improve the $\ell_2$ performance of $\ell_\infty$ image codecs such as near-lossless CALIC, Zhou \textit{et al.} \cite{zhou2012ell_} proposed a soft decoding process.  It is formulated as an image restoration task and the solution is based on the context modeling of the quantization errors while maintaining a strict error-bound.  Chuah \textit{et al.} replaced the uniform quantizer of near-lossless CALIC by a set of context-based $\ell_2$-optimized quantizers \cite{chuah2013ell_}, with a goal of minimizing $\ell_2$ distortions and entropy while remaining within a strict error bound. Li \textit{et al.} \cite{li2014sparsity} improved the performance through sparsity-driven restoration utilizing a combination of $\ell_1$ and $\ell_2$ losses while restricting the range to a tight error bound. Zhang \textit{et al.} \cite{zhang2020ultra} utilized a soft-decompression network (SDNet) to improve the $\ell_2$ performance based on a learning model of natural images through the encoder-decoder network architecture while restricting the range of the output using truncated activation.

Recently, Bai \textit{et al.} \cite{bai2021learning} proposed a pure-CNN method of $\ell_\infty$-constrained near-lossless image compression by jointly learning lossy image and residual compression.

\subsection{Light field restoration}
\label{sec:lfrestoration}
\noindent
Compression artifacts removal is a special case of restoration.
Apart from compression there are other degradation causes in light fields generated using view synthesis operation, including noises, insufficient sampling, etc. In \cite{srinivasan2017learning}, an intermediate light field is synthesized using a single image by disparity map learning for a specific type of content i.e. flowers. The intermediate light field is further refined by utilizing a 3D CNN to model occlusion and non-Lambertian surfaces. In \cite{wu2017light, wu2018light}, a detail restoration network is used to restore high-frequency details in the light field using epipolar plane image representation. Yeung \textit{et al.} \cite{yeung2018fast} uses a two-stage process, namely view synthesis and view refinement networks, for light field reconstruction using four corner views. To efficiently exploit 4D light field structures a sequence of Spatial-Angular Separable (SAS) convolutions is used in the view synthesis network, whereas a set of 4D convolutions is used to generate a refined light field. In \cite{jin2020learning}, Jin \textit{et al.} incorporated SAS convolution to improve the refinement process of a view synthesis network. In \cite{gul2020light}, Gul \textit{et al.} utilizes an attention-based network (CBAM) to precisely estimate occluded and reflective features in the light field refinement process.

\begin{figure*}[!t]
\centering
\includegraphics[width=0.9\textwidth]{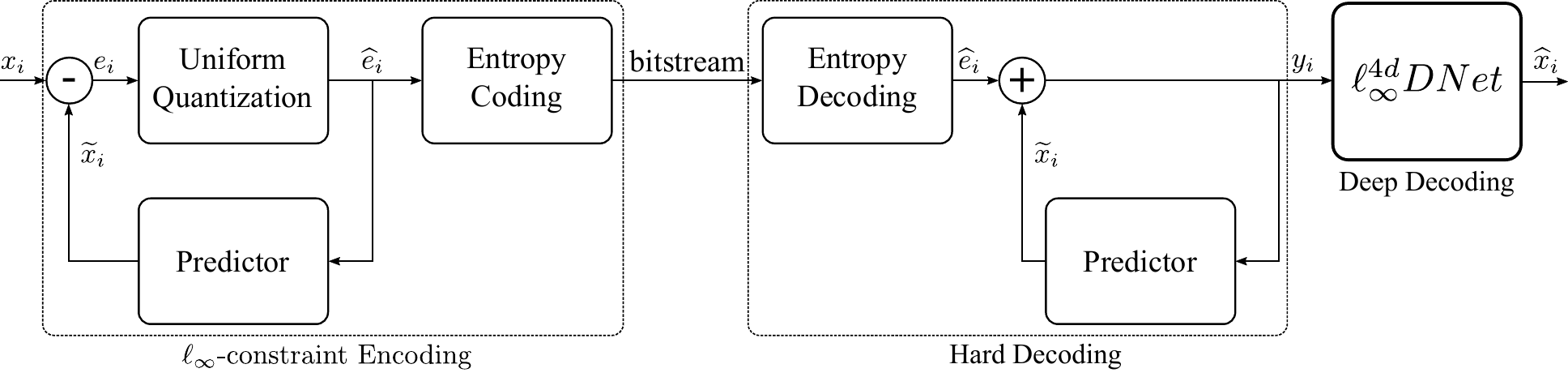}
\caption{\label{fig:overall_blockdiag}Overall architecture of the proposed $\ell_\infty^{4d}$DNet based compression system incorporating the proposed deep decompression network $\ell_\infty^{4d}$DNet.}
\vspace{-1.5em}
\end{figure*}

The task of light field super-resolution can also be viewed as a way to restore a high-resolution light field by incorporating high-frequency details that are removed due to down-sampling. In \cite{fan2017two}, the high-resolution light field is obtained in two stages. The first step individually upsamples each view based on the natural image property using a very deep CNN. The second stage exploits inter-view correlation to further improve the resolution. SAS convolution is used again in \cite{yeung2018light}, this time to improve light field spatial resolution.

In summary a few common choices of network architectures to model light field structures are published, i.e. SAS or 4D convolutions.
However, these architectures are unable to accurately model the light field structure either due to insufficient information flow between the low dimensional spaces as in SAS or inadequate network depth required to efficiently exploit the 4D light field structure using 4D convolutions.
For this reason, we propose an alternative network architecture $\ell_\infty^{4d}$DNet, which can efficiently model the overall light field structure relying on the learned 2D models of low-dimensional light field representations.
In this context, the learned models of these representations independently detect the compression artifacts in the light field structure. Based on the reliability of a 2D model for a local 4D region/pixel, the weighting network in $\ell_\infty^{4d}$DNet assigns the contribution of these models to compensate the compression artifacts.

\section{System overview}
\label{sec:overview}
\subsection{Light field notation}
\label{sec:lfoverview}
\begin{figure}[t!]
\centering
\includegraphics[width=0.4\textwidth]{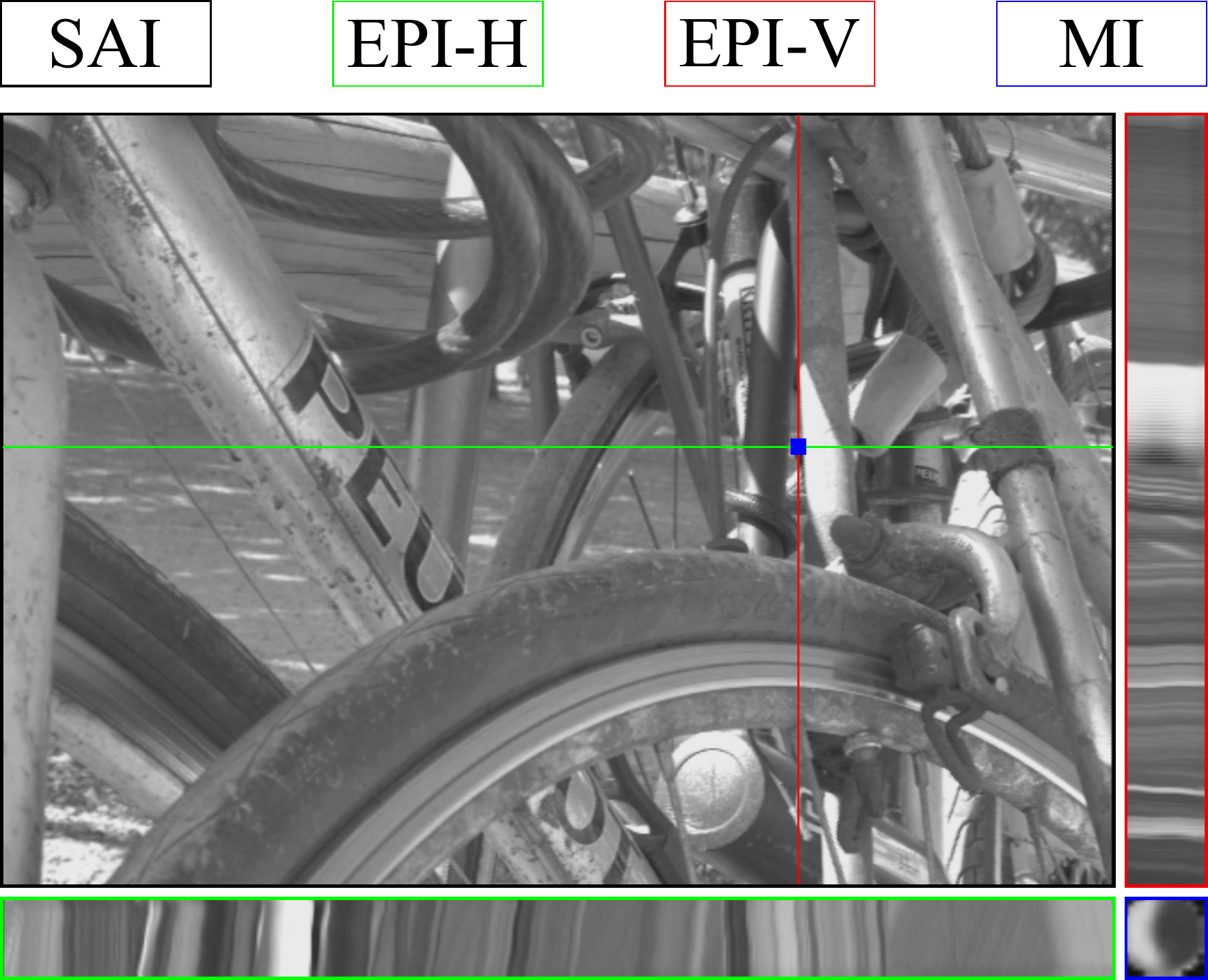}
\caption{\label{fig:lf_repr} Four distinct light field representations. Each of them is a 2D subspace of the four dimensions of light field.}
\vspace{-1.0em}
\end{figure}

\noindent
Light field $\mathcal{L}\left(s,t,u,v\right)$ is a 4D dataset composed of two orthogonal components for each of the angular ($s, t$) and the spatial ($u, v$) dimensions, where $s$ and $u$ being the vertical components, while $t$ and $v$ are the horizontal components of these dimensions. Grouping two sub-spaces of the 4D dimensions in different combinations creates four different light field representations, as shown in Fig.~\ref{fig:lf_repr}.
% A particular perspective of a light field can be selected by indexing angular dimensions ($s, t$) (formally known as Sub-Aperture Image (SAI)).
A regular 2D image representation, commonly known as Sub-Aperture Image (SAI), is originated by indexing angular dimensions ($s, t$).
Another representation is Epipolar Plane Image (EPI), which is formed by gathering parallel components of spatial and angular dimensions. For instance, a vertical-EPI (EPI-V) representation can be obtained by fixing the angular coordinate $t$ and the spatial coordinate $v$; symmetrically,
a horizontal-EPI (EPI-H) representation can be obtained by fixing the angular coordinate $s$ and the spatial coordinate $u$.  Finally, a Micro Image (MI) representation can be achieved by fixing the spatial coordinates $u$ and $v$, which contains the angular information at spatial location $(u,v)$.

\subsection{Lossless EPI compression of light field}
\noindent
In \cite{mukati2021improved}, Mukati \textit{et al.} proposed a context-adaptive scheme for lossless EPI compression of light field (EPIC).  Lossless EPIC is an extension of CALIC for lossless
compression of light field images. It exploits the correlation present in the 4D light field structure for improved prediction and context formation.
This compression system was mainly designed for thin-baseline light field such as the ones obtained using plenoptic light field cameras, which results in an EPI representation composed of linear continuous textures.
The EPIC encoder exploits the simplicity of this representation for effective prediction and context formation.

\subsection{$\ell_\infty^{4d}$DNet based light field compression}
\label{sec:linfepic}
\noindent
To pursue higher compression ratio while maintaining high data fidelity, we propose the $\ell_\infty^{4d}$DNet based light field compression system, which is schematically described in
Fig.~\ref{fig:overall_blockdiag}. We modify the basic structure of lossless EPIC to operate in near-lossless mode for higher compression gains and add the deep CNN decoder $\ell_\infty^{4d}$DNet for higher quality light field reconstruction.
The details and rationale of the $\ell_\infty^{4d}$DNet based compression system are presented in this section.

The encoding scheme in this system is a $\ell_\infty$-constrained EPIC encoder, which represents a class of DPCM encoders. It sequentially encodes the pixels of the given light field $X$ by progressively traversing along all four dimensions. For the intensity $x_i$ of each pixel $i$, $\forall i | i \in \lbrace s,t,u,v\rbrace$, the predictor returns an estimate $\tilde{x}_i$ based on the neighboring intensities and statistics of the causal neighborhood. Eventually, the resulting prediction error $e_i$ can be sequentially encoded using an entropy coding scheme for lossless encoding.
At this point, the achievable compression ratio is bounded by the entropy of the prediction error. Thus to achieve higher compression gains, we reduce the entropy of the prediction error by quantizing the prediction error. Uniform quantization of the prediction error guarantees that the absolute reconstruction error remains under a certain maximum value $\tau$. This form of compression is called $\ell_\infty$-constrained compression and it offers high-fidelity data reconstruction, when the $\tau$ is restricted to some small value.
To balance the tradeoff between compression ratio and data fidelity, the quantization parameter $\tau$ is used to quantize the prediction error $e$ as follows:
\begin{equation}
\label{eq:quant}
    \hat{e} = Q\left[e\right] = \left\lfloor\frac{e+\tau}{2\tau+1}\right\rfloor.
\end{equation}
% \noindent
Hence, instead of the actual error $e$, the quantized prediction error $\hat{e}$ is transmitted to the decoder through entropy coding.

\begin{figure*}[!t]
     \centering
     \begin{subfigure}[b]{0.455\textwidth}
         \centering
         \includegraphics[width=\textwidth]{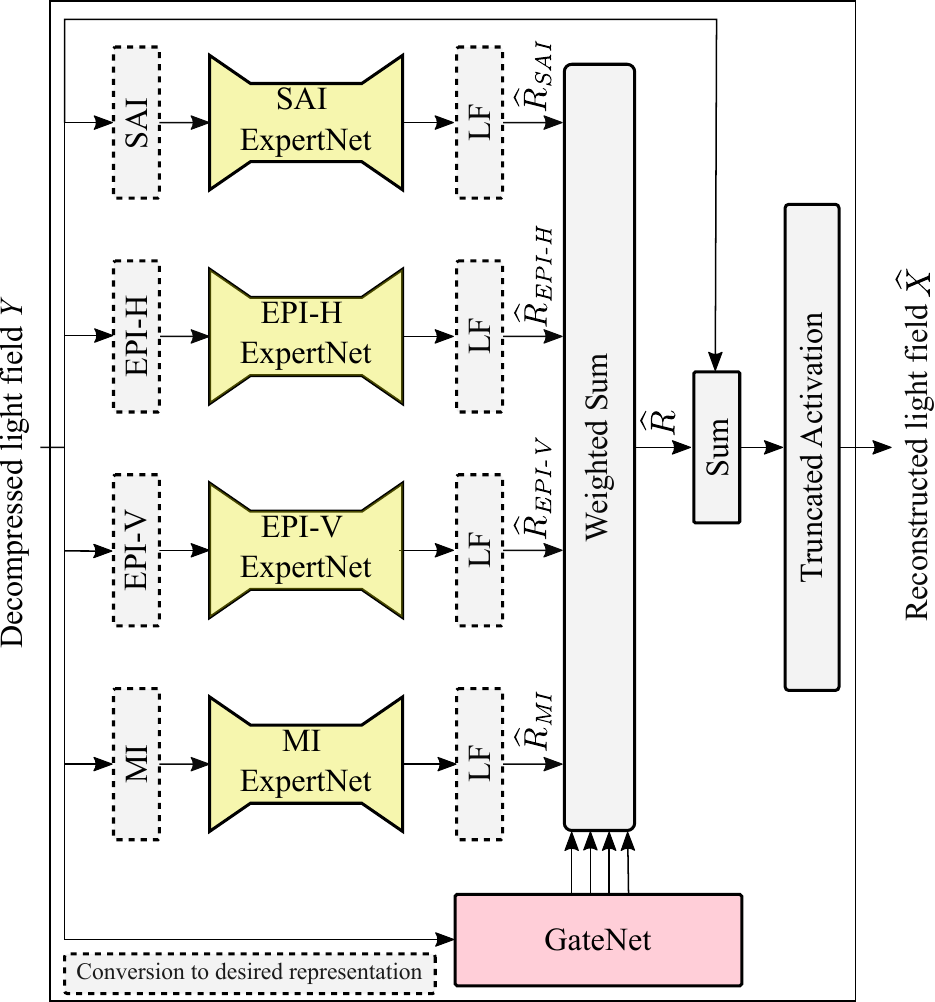}
         \\
         \footnotesize{(a) $\ell_\infty^{4d}$DNet}
     \end{subfigure}
     \hfill
     \begin{subfigure}[b]{0.36\textwidth}
         \centering
         \includegraphics[width=\textwidth]{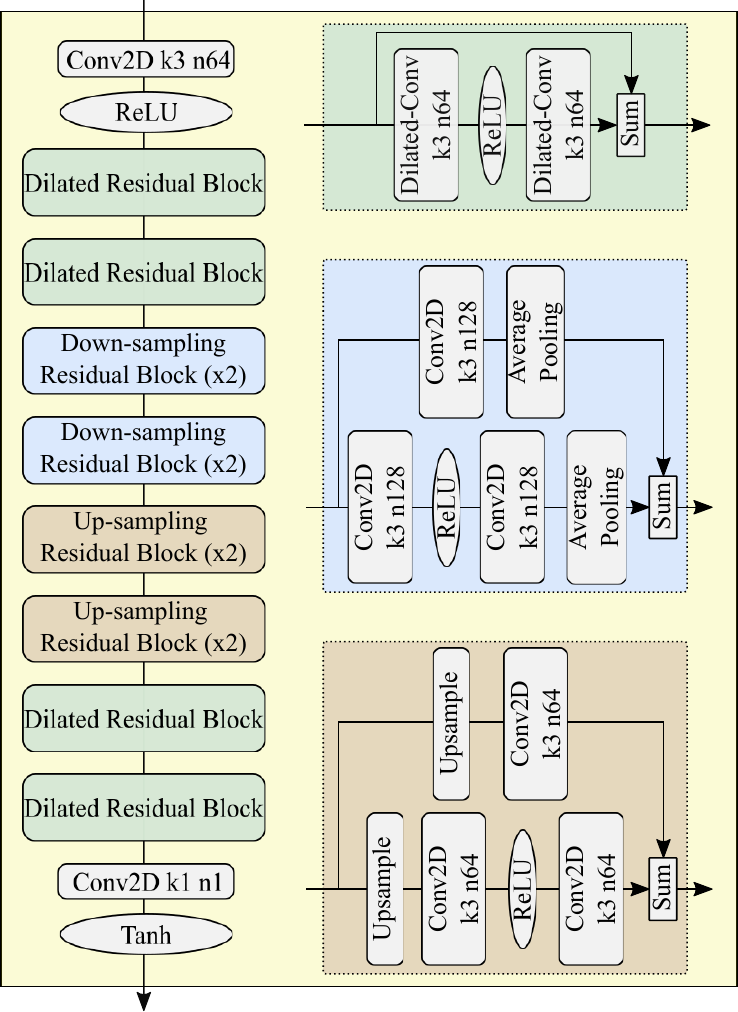}
         \\
         \footnotesize{(b) ExpertNet}
     \end{subfigure}
     \hfill
     \begin{subfigure}[b]{0.14\textwidth}
         \centering
         \includegraphics[width=\textwidth]{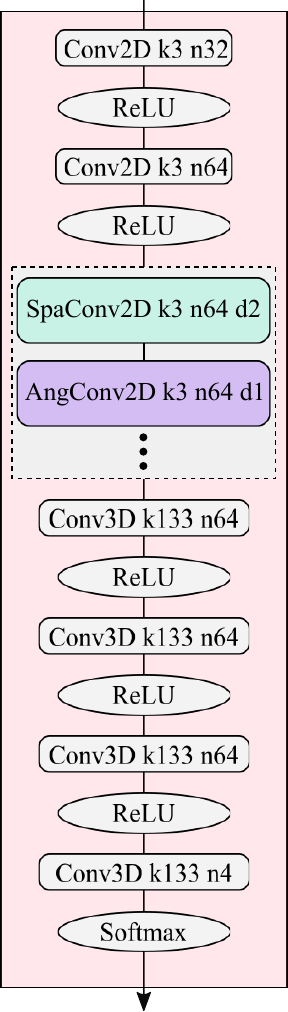}
         \\
         \footnotesize{(c) GateNet}
     \end{subfigure}
    \caption{\label{fig:soft_blockdiag}Blockdiagram of the proposed network, $\ell_\infty^{4d}$DNet.}
    \vspace{-1.5em}
\end{figure*}

Meanwhile, at the decoder, the hard EPIC decoder initially decompresses an $\ell_\infty$-constrained light field $Y$ by reversing the encoding pipeline excluding the quantization block. Precisely, the decoder sequentially decompresses intensity of a pixel $y_i$, by adding the quantized prediction error $\hat{e}_i$ from the entropy decoder to the prediction $\tilde{x}_i$.
Thus, the error $\tilde{e}_i = y_i - x_i, \forall i,$ in the resulting light field $Y$ is bounded by the range defined by the quantization parameter $\tau$, i.e. $\lvert\tilde{e}_i\rvert \leq \tau$ (Eq. \ref{eq:quant});
where $x_i$ and $y_i$ represent the intensities of the pixel $i$ of the original light field $X$  and the light field $Y$ decompressed by the EPIC decoder, respectively.

To enhance the quality of the decompressed light field $Y$, we incorporate the proposed $\ell_\infty^{4d}$DNet in the proposed compression system to perform deep decompression of the latent light field.
The $\ell_\infty^{4d}$DNet in presence of well-defined light field model utilizes the guaranteed error bound $\tau$ in the decompressed light field $Y$ as a prior information to decompress $\ell_\infty$-constrained light field $\widehat{X}$, having superior $\ell_2$ performance.

\begin{equation*}
    \widehat{X} = f_{\ell_\infty^{4d}}\left(Y\right)
\end{equation*}

The additional novelty of this paper is the use of deep decompression network $\ell_\infty^{4d}$DNet in conjunction with the EPIC codec, aiming at ultra-high fidelity reconstruction of the light field.  This enhances the $\ell_2$ quality of the hard decompressed light field $Y$ without requiring any additional information from the encoder.
In the next section, we will explain the network architecture of the proposed $\ell_\infty^{4d}$DNet.

\section{Design of $\ell_\infty^{4d}$DNet}
\label{sec:slfdnet}
\noindent
The design of the $\ell_\infty^{4d}$DNet aims to build a light field restoration model to remove the compression artifacts in the decompressed light field $Y$.
For its realization, a straightforward way of exploiting signal redundancies in the four dimensions of the light field is to use 4D convolutional kernels that extract 4D features.
However, such a 4D approach will require large amount of computational resources.
Recall from Section \ref{sec:lfoverview} and Fig.~\ref{fig:lf_repr}) that the 4D light field has four 2D representations, SAI, EPI-H, EPI-V, and MI.  To make $\ell_\infty^{4d}$DNet computationally feasible, it is designed to operate on these four 2D light field representations, instead.

% Explanation of the proposed design
Intelligently fusing light field estimates from multiple representations can result in a robust light field reconstruction.  Different 2D light field representations offer information on different aspects of the light field.  They have their strengths and weaknesses depending on scene compositions and hence need to be combined properly to aid the reconstruction process.
Therefore for $\ell_\infty^{4d}$DNet, we propose a mixture-of-experts like architecture, as shown in Fig. \ref{fig:soft_blockdiag}a, which utilizes a gating network (GateNet) to weigh out the contributions of the estimates from the four expert networks (ExpertNet(s)), where each ExpertNet uniquely processes the light field in one of the four representations described in Section \ref{sec:lfoverview}. Training the $\ell_\infty^{4d}$DNet in end-to-end fashion, forces each of the four ExpertNets to model the corresponding light field representation. The detailed architecture of the proposed $\ell_\infty^{4d}$DNet is presented in the following sub-section.

\subsection{Network architecture}
\label{sec:networkarchitecture}
\noindent
The proposed $\ell_\infty^{4d}$DNet operates as follows.
To reconstruct a higher quality light field, $\ell_\infty^{4d}$DNet estimates a residual light field $\widehat{R}$ to compensate errors in the decompressed light field $Y$, instead of estimating the absolute intensities.
For this purpose, each of the four ExpertNets $f_{E,k}\left(\boldsymbol{\cdot}\right)$ estimates a residual light field $\widehat{R}_k$ by processing the corresponding representation $k$ of the decompressed light field $Y$.
\begin{equation*}
\begin{array}{lcr}
&\widehat{R}_k = f_{E,k}\left(\mathcal{T}_k\left(Y\right)\right), & k \in K
\end{array}
\end{equation*}
\noindent
where $K=\lbrace \text{SAI}, \text{EPI-H}, \text{EPI-V}, \text{MI}\rbrace$, whereas,
$\mathcal{T}_k\left(\boldsymbol{\cdot}\right)$ reshapes the light field to the subsequent $k^{th}$ representation.
The decompressed 4D light field $Y$ is directly passed to the GateNet ($f_{G}\left(\boldsymbol{\cdot}\right)$) to determine pixel-wise weights $\textbf{W}, \textbf{W} \in 	\mathbb{R}^{S{\times}T{\times}U{\times}V{\times}4}$, for each of the estimated residuals.
\begin{align*}
    \textbf{W} = \left(W_{\text{SAI}}, W_{\text{EPI-H}}, W_{\text{EPI-V}}, W_{\text{MI}}\right) = f_{G}\left(Y\right)
\end{align*}

Finally, the estimated weights $W_K$ are used to average the residual light fields $\widehat{R}_K$, to obtain a robust estimate of a residual light field $\widehat{R}$. The residual $\widehat{R}$ is eventually used to compensate errors in the decompressed light field $Y$, as follows:
\begin{equation*}
    \widehat{X} = \phi \left(Y + \sum_k W_k \odot \widehat{R}_k\right)
\end{equation*}
\noindent
where, $\phi\left(\boldsymbol{\cdot}\right)$ is the truncated activation function. As described in Section \ref{sec:linfepic}, the reconstruction error in the decompressed light field $Y$ is bounded by the range $\left[-\tau,\tau\right]$. The ability to wholly compensate for the reconstruction error in $Y$, is achieved by extending the range of the estimated residuals to $\left[-\tau,\tau\right]$. Similar to \cite{zhang2020ultra}, the residual range is practically restricted by applying the truncated activation function to the sum of the residual light field $\widehat{R}$ and the decompressed light field $Y$.
Using the estimated residual $\widehat{R}$ to compensate reconstruction errors in the decompressed light field $Y$, considering that both having a range of $\left[-\tau,\tau\right]$, will result in a restored light field $\widehat{X}$ with a possible reconstruction error within the range $\left[-2\tau,2\tau\right]$.

\subsubsection{ExpertNet(s)}
Each ExpertNet is provided with a unique two-dimensional light field representation of the four representations and estimates a residual light field based on inconsistencies detected in the corresponding representation.
Therefore, individual light field representations are formed by reshaping the decompressed light field $Y$ to be passed as an input to ExpertNets.
As described in Section \ref{sec:lfoverview}, a particular light field representation is formed by extracting corresponding two dimensions of the four dimensions of the light field. The product of the remaining two dimensions serves as the number of the particular representations available out of the light field. Thus, a light field with dimensions $S,T,U,V$, yields $S\times T$ SAIs, $S\times U$ EPI-Hs, $T\times V$ EPI-Vs, and $U\times V$ MIs.

To model each representation by the respective ExpertNet, an encoder-decoder architecture is used. Specifically, we choose the $\ell_\infty$-SDNet architecture of \cite{zhang2020ultra}, with a slight modification based on the modeling requirement of each representation. The detailed network architecture of ExpertNet is presented in Fig. \ref{fig:soft_blockdiag}b.
Each ExpertNet starts by extracting some low-level features from the current representation using a $3\times 3$ 2D convolution, followed by a series of 8 residual blocks. Finally, a single channel output is generated using a $1\times 1$ 2D convolution.
The network design can efficiently model each representation in the presence of the strictly bounded error as a strong prior. The receptive field of the network is widened through down-sampling and up-sampling operations, and dilated convolutions in the residual blocks.

Out of the 8 residual blocks, there are 4 dilated, 2 down-sampling, and 2 up-sampling residual blocks. The Dilated Residual Block keeps the spatial resolution intact, while it is responsible for collecting features from a larger receptive field. The Dilated Residual Block consists of two $3\times 3$ dilated convolutions (dilation parameter $= 2$) with a skip connection. The Down-sampling Residual Block finds a sparser representation of the features with twice the channel dimension and half the spatial dimensions, whereas the Up-sampling Residual Block reconstructs the information through the sparse representation provided by the down-sampling residual block. Each down-sampling and up-sampling residual block utilizes three $3\times 3$ convolutions, whereas average pooling and upsampling functions are used to scale the spatial dimensions.
Since the three representations, EPI-H, EPI-V, and MI, are highly predictive, they require fewer network parameters for their modeling, hence, the corresponding ExpertNets deliberately utilize half the number of channels compared to the one used for the SAI ExpertNet; eventually halving the size of the rest of the ExpertNets. Furthermore, the angular dimensions are typically much smaller compared to the spatial dimensions in the light field, and usually do not require a large receptive field to capture the redundancy in its structure, therefore the down-sampling and up-sampling operators do not scale the angular dimensions in the concerned ExpertNets. The detailed network block diagram is provided in Fig. \ref{fig:soft_blockdiag}b.

\begin{figure}[t!]
\centering
\includegraphics[width=0.35\textwidth]{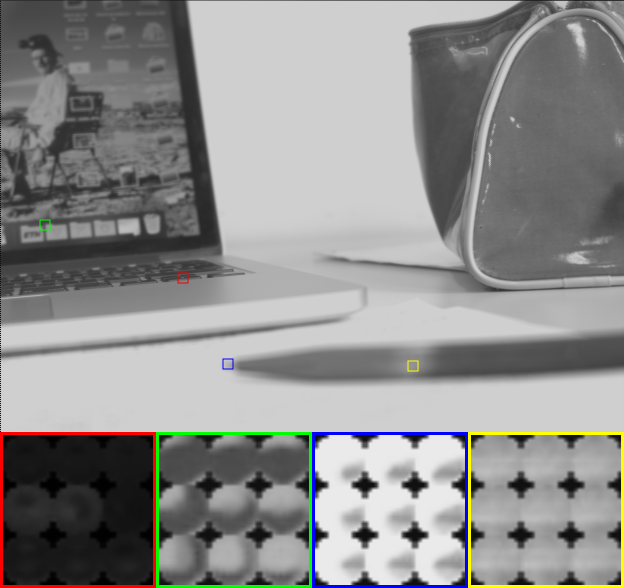}
\caption{\label{fig:mi_regions} Zoomed patches of the MI regions to demonstrate the simplicity of this representation. The patch with red boundary highlights a focused region, which is expected to have the same intensity across the MI. The green and blue patches represents textures in the unfocused regions, representing MI with simple textures. The yellow patch illustrates a texture-less unfocused region, which also results in a flat MI surface like the red patch.}
\end{figure}

\subsubsection{GateNet}
\label{sec:gatenet}
% 1. Responsibility of gating network.
% a. Show why it is easy to learn when it is focusing using macro
The GateNet aims to determine the weights for the four expert residuals such that the average residual light field can be used for robust and optimal quality restoration of light field. The GateNet should be able to locally assign a larger weight to an expert residual likely to produce a higher quality. The network should therefore learn the local effectiveness of each light field representation towards light field restoration under different challenging conditions.
More precisely, the overall reconstruction quality is derived by the following characteristics of the light field representations.

\textit{Micro Images:} Due to the thin-baseline of the views in plenoptic cameras, the range of the disparity among the adjacent views is limited to only a few pixels. As MI representation is a collection of intensities from the same spatial indices from all the views, the textures culminating in this representation are virtually smooth as shown in Fig. \ref{fig:mi_regions}.
In majority cases, the intensities in an MI are overall constant. For example, if the MI is captured for a region with zero disparity. On the other hand, the capture of a smooth surface will also result in a flat MI.
Additionally, in a few cases, an MI can be observed as a zoomed-in snapshot of a small region. In this case, the representation entails the properties of natural images.

\textit{Epipolar Plane Images:}
As shown in Fig. \ref{fig:lf_repr}, again due to thin-baseline of light field views, the EPIs consist of linear edges and textures, where the slope of a line determines the depth of an object in the scene. Therefore, it is possible to obtain a high-quality restoration regardless of a feature's depth. However, the restoration process in presence of occlusion and reflection becomes challenging. The linear nature of an EPI becomes distorted, for instance in the region of an EPI-H capturing a feature occluded by a vertical edge of a foreground object, or in the region of an EPI-V capturing a feature occluded by a horizontal edge of a foreground object.
However, the presence of vertical occlusion does not distort the linearity of EPI-V and thus can be used for effective restoration. Similar is true for EPI-H in presence of horizontal occlusion.

\textit{Sub-Aperture Images:} SAI possesses natural image properties. Natural images are distinctive from the large possible permutations of an image as they contain particular types of structure. For robust restoration in challenging cases, this representation can serve as a dependable source for prediction.

The network design of GateNet should be able to exploit 4D correlation to learn the features of each of the representations as described above.
The GateNet will be provided with a 4D light field of size $S\times T\times U\times V$ at its input, whereas, it is expected to return weights for all of the four estimated residuals, such that the final output size is $S\times T\times U\times V\times 4$.
To maintain a relatively low complexity of the proposed $\ell_\infty^{4d}$DNet, instead of using 4D CNNs, we considered alternate network architecture for GateNet, which can exploit redundancy in the light field with relatively lower complexity.
The Spatial Angular Separable (SAS) convolutions described in Section \ref{sec:relevant}, is used in the context of light field super-resolution and light field view synthesis.
A sequence of SAS convolutions provides a computationally- and memory-efficient alternative for extraction of spatial-angular joint features of the light field, sufficiently serving as an approximation of 4D convolution.
In this project, we use the SAS convolution in GateNet for the purpose of estimating weights to combine expert residuals.
The design of the network is highly inspired by the ``light field blending module'' of \cite{jin2020learning}, which incorporates a sequence of SAS convolutions.
The details of the adopted network design for GateNet are shown in Fig. \ref{fig:soft_blockdiag}c.
The light field $Y$ presented to the input of the GateNet is initially reshaped to $S\times T$ SAI representations, to extract 2D spatial features using a sequence of two 2D convolutions. Afterward, the signal is passed through a sequence of three SAS convolutional operators. Each iteration of SAS sequentially applies 2D convolution, to the spatial representation, i.e. SAI, and to angular representation, i.e. MI, involving the required reshaping process. Thereafter, the intermediate signal is converted to a 3D representation by vectorizing the two angular dimensions, to be passed to a sequence of four 3D convolutions. The last convolution block generates a four channel signal in relation to the four expert estimates. Except for the last convolution, each of the convolutions is followed by a ReLU activation function. The signal is finally passed through a Softmax activation function such that the output of the four channels sums up to one.

\section{Experimental Results}
\label{sec:analysis}
\noindent
In this section, we evaluate the performance of the $\ell_\infty^{4d}$DNet based compression system and compare it with existing light field compression methods.
\begin{figure*}[!th]
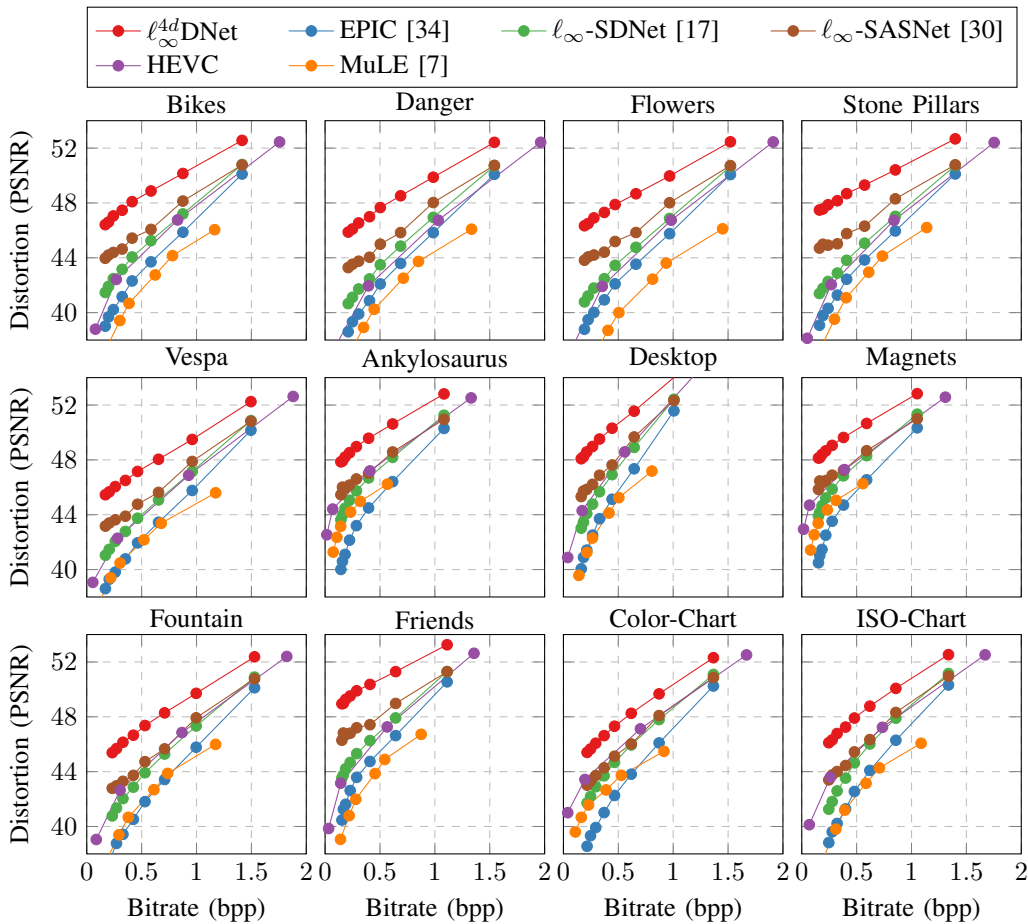

 \centering
\include{customplots/rate_psnr}
\vspace{-2em}
\caption{\label{fig:rd_residual}RD performance comparison of the tested light field coding schemes on the evaluation set. The anchors for HEVC are generated by setting QP = $\lbrace6,12,18,24\rbrace$, whereas, for MuLE $\lambda = \lbrace3,6,11,22,43,86\rbrace$ were used. For EPIC, $\ell_\infty$-SDNet, $\ell_\infty$-SASNet, and $\ell_\infty^{4d}$DNet system, $\tau=\lbrace1..8\rbrace$ were used at the encoder to generate data points.}
\vspace{-1.0em}
\end{figure*}

\subsection{Dataset and experiment setup}
\label{sec:analysis:dataset}
\noindent
For training and performance evaluation, we use the EPFL light field dataset \cite{rerabek2016new}, downloaded from \cite{plenodb}.
The dataset is composed of $118$ thin-baseline light field images captured using Lytro Illum camera at $10$-bit depth. The light fields in the dataset are categorized into $10$ different categories based on the content of the scene. Each light field in the dataset has $15\times 15$ views of $434\times625$ spatial resolution in $3$ RGB channels. We prepare the dataset for training and evaluation as follows.
Raw light fields are gamma corrected ($\gamma = 0.45$), converted to Y-channel, and quantized to $8$-bit precision.
The resulting light fields serve as the undistorted ground truth, out of which $50$ light fields are set aside for training, and $12$ light fields are used for evaluating all the tested methods in the subsequent sub-sections.
The light fields in our training and evaluation datasets are compressed by the $\ell_\infty$-constrained EPIC codec with $\ell_\infty$ bound $\tau$ ranging from $1$ to $8$. The corresponding hard decompressed light fields are fed to $\ell_\infty^{4d}$DNet to be further restored or soft decoded.
The light fields in the evaluation dataset are the $12$ light fields selected for evaluation in the light field compression challenge \cite{rerabek2016icme}.
For the training set, an equal number of light fields are picked from each of the $10$ categories of scenes, to avoid model overfitting.

To train the proposed $\ell_\infty^{4d}$DNet, patches of $64\times 64$ pixels are extracted randomly from the light fields in the training dataset, whereas the simple $\ell_2$ loss is used for network optimization. The network is trained using Adam optimizer with default beta parameters, i.e. 0.9 and 0.999. It is trained for $600$ epochs, with the initial $400$ epochs of $1e-4$ learning-rate, followed by $200$ epochs of $1e-5$ learning-rate for fine-tuning. The training is carried out on the PyTorch framework running on DTU HPC LSF10 cluster with Tesla A100-PCIE GPU.

To evaluate the performance of the proposed $\ell_\infty^{4d}$DNet based compression system, we report the experimental results and compare them with those of other light field compression methods in the literature.
We use HEVC \cite{sullivan2012overview} as a reference benchmark to compare the performances of all tested methods.
JPEG Pleno provides two standard light field coding solutions, i.e. MuLE (based on 4D transformation) and WaSP (based on 4D prediction) \cite{schelkens2019jpeg}. For thin-baseline light fields, MuLE is superior to WaSP due to its ability to decorrelate a small 4D neighborhood by confining the energy of the light field to only a few 4D transform coefficients \cite{alves2020jpeg}.
Therefore, we exclude WaSP and include MuLE in the group of testing methods.

\begin{table*}[!t]
\centering
\caption{Average BD-PSNR and BD-Rate gains over HEVC on the evaluation set.}
\begin{tabularx}{0.97\textwidth}{lSSSSSSSSSS}
\toprule
 & \multicolumn{2}{c}{$\ell_\infty^{4d}$DNet} & \multicolumn{2}{c}{EPIC \cite{mukati2020epipolar}} & \multicolumn{2}{c}{$\ell_\infty$-SDNet \cite{zhang2020ultra}}  & \multicolumn{2}{c}{$\ell_\infty$-SASNet \cite{jin2020learning}}  & \multicolumn{2}{c}{MuLE \cite{schelkens2019jpeg}}\\
\cmidrule(lr){2-3} \cmidrule(lr){4-5} \cmidrule(lr){6-7} \cmidrule(lr){8-9} \cmidrule(lr){10-11}
& \multicolumn{1}{c}{{BD-PSNR}} & \multicolumn{1}{c}{BD-Rate} & \multicolumn{1}{c}{{BD-PSNR}} & \multicolumn{1}{c}{BD-Rate} & \multicolumn{1}{c}{{BD-PSNR}} & \multicolumn{1}{c}{BD-Rate} & \multicolumn{1}{c}{{BD-PSNR}} & \multicolumn{1}{c}{BD-Rate} & \multicolumn{1}{c}{{BD-PSNR}} & \multicolumn{1}{c}{BD-Rate}\\
Sequence & \multicolumn{1}{c}{[\text{dB}]} & \multicolumn{1}{c}{($\%$)} & \multicolumn{1}{c}{[\text{dB}]} & \multicolumn{1}{c}{($\%$)} & \multicolumn{1}{c}{[\text{dB}]} & \multicolumn{1}{c}{($\%$)} & \multicolumn{1}{c}{[\text{dB}]} & \multicolumn{1}{c}{($\%$)} & \multicolumn{1}{c}{[\text{dB}]} & \multicolumn{1}{c}{($\%$)}\\
%FLIF
\midrule
Bikes	        & 3.90 & -49.6 & -1.44 & 33.5 &  0.23 &  -4.5 & 1.46 & -22.6 &-2.89 &  93.7\\
Danger 	        & 4.40 & -47.3 & -0.79 & 14.5 &  0.56 & -10.0 & 1.99 & -25.7 &-2.28 &  53.7\\
Flowers         & 4.50 & -47.8 & -0.79 & 14.8 &  0.38 &  -5.6 & 2.05 & -24.9 &-3.17 &  75.5\\
Stone Pillars   & 4.77 & -51.9 & -1.15 & 34.3 &  0.41 &  -8.9 & 2.14 & -26.6 &-2.52 &  75.0\\
Vespa           & 3.17 & -42.2 & -1.47 & 28.3 &  0.12 &  -0.5 & 1.02 & -16.6 &-1.97 &  69.5\\
Ankylosaurus    & 2.24 & -49.9 & -2.81 & 92.2 & -0.56 &  19.8 & 0.03 &  -5.1 &-1.94 & 173.7\\
Desktop  	    & 2.79 & -38.7 & -2.41 & 47.6 & -0.53 &  11.8 & 0.43 &  -8.5 &-3.16 & 111.1\\
Magnets 	    & 2.18 & -49.2 & -2.64 & 89.4 & -0.52 &  18.2 & 0.10 &  -7.7 &-2.05 & 194.7\\
Fountain        & 2.58 & -36.5 & -2.34 & 39.5 & -0.48 &   9.0 & 0.18 &  -2.9 &-2.41 &  69.0\\
Friends         & 4.15 & -57.7 & -1.30 & 33.1 &  0.31 &  -6.0 & 1.63 & -29.6 &-2.78 & 102.7\\
Color-Chart  	& 1.60 & -31.4 & -2.96 & 56.8 & -0.77 &  15.5 &-0.42 &  10.0 &-2.28 & 123.0\\
ISO-Chart 	    & 2.24 & -36.7 & -2.49 & 40.5 & -0.64 &   8.4 &-0.02 &  -0.5 &-3.35 & 108.2\\
\midrule
Average         & 3.21 & -44.9 & -1.88 & 43.7 & -0.12 &  3.9  & 0.88 & -13.4 &-2.66 & 104.2\\
\bottomrule
\end{tabularx}
\label{tab:bjontegaard}%
\end{table*}

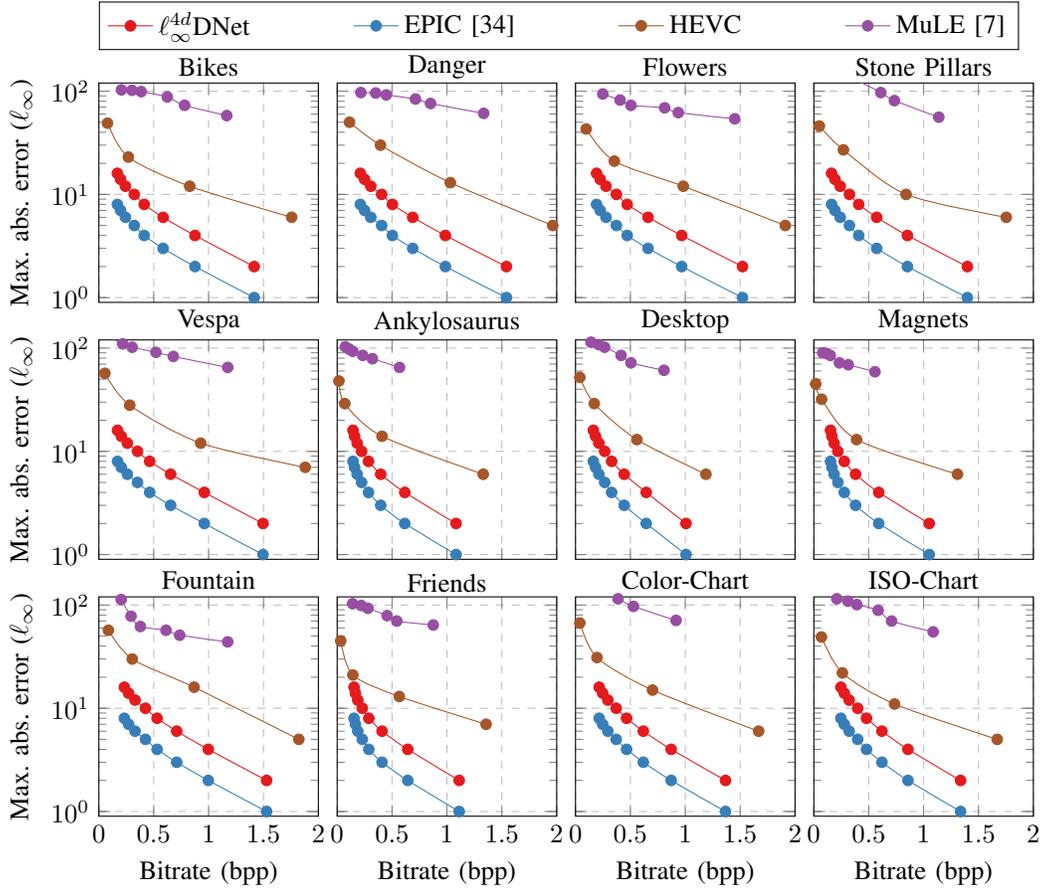
\begin{figure*}[!h]
 \centering
\begin{tikzpicture}
	\begin{groupplot}[
		group style={group size=4 by 3,
			x descriptions at=edge bottom,
			y descriptions at=edge left},
		my style1,
		xmin=0,xmax=2,
        ymin=0.9,ymax=120,
        xtick={0, 0.5, 1.0, 1.5, 2.0},
        ytick={1, 10, 100},
		legend pos=north west,
		ymajorgrids=true,
		xmajorgrids=true,
		grid style=dashed,
		ymode=log,
	%log ticks with fixed point,
	% for log axes, x filter operates on LOGS.
	% and log(x * 1000) = log(x) + log(1000):
	%x filter/.code=\pgfmathparse{#1 + 6.90775527898214},
	]
	\nextgroupplot[residual legend style1, title style={yshift=-1.5ex,}, title={Bikes}]
		\addplot[smooth, color=clr_plt1, mark=*]
			table [x=mhvs_rate, y=mhvs_pae, col sep=comma] from {csv/Urban_Bikes.csv};
		\addplot[smooth, color=clr_plt2, mark=*]
			table [x=epic_rate, y=epic_pae, col sep=comma] from {csv/Urban_Bikes.csv};
		\addplot[smooth, color=clr_plt4, mark=*]
			table [x=hevc-inter_rate, y=hevc-inter_pae, col sep=comma] from {csv/Urban_Bikes.csv};
		\addplot[smooth, color=clr_plt5, mark=*]
			table [x=mule_rate, y=mule_pae, col sep=comma] from {csv/Urban_Bikes.csv};
	\nextgroupplot[cycle list shift=1, title style={yshift=-1.7ex,}, title={Danger}]
		\addplot[smooth, color=clr_plt1, mark=*]
			table [x=mhvs_rate, y=mhvs_pae, col sep=comma] from {csv/Grids_Danger_de_Mort.csv};
		\addplot[smooth, color=clr_plt2, mark=*]
			table [x=epic_rate, y=epic_pae, col sep=comma] from {csv/Grids_Danger_de_Mort.csv};
		\addplot[smooth, color=clr_plt4, mark=*]
			table [x=hevc-inter_rate, y=hevc-inter_pae, col sep=comma] from {csv/Grids_Danger_de_Mort.csv};
		\addplot[smooth, color=clr_plt5, mark=*]
			table [x=mule_rate, y=mule_pae, col sep=comma] from {csv/Grids_Danger_de_Mort.csv};
	\nextgroupplot[cycle list shift=2, title style={yshift=-1.5ex,}, title={Flowers}]
		\addplot[smooth, color=clr_plt1, mark=*]
			table [x=mhvs_rate, y=mhvs_pae, col sep=comma] from {csv/Nature_Flowers.csv};
		\addplot[smooth, color=clr_plt2, mark=*]
			table [x=epic_rate, y=epic_pae, col sep=comma] from {csv/Nature_Flowers.csv};
		\addplot[smooth, color=clr_plt4, mark=*]
			table [x=hevc-inter_rate, y=hevc-inter_pae, col sep=comma] from {csv/Nature_Flowers.csv};
		\addplot[smooth, color=clr_plt5, mark=*]
			table [x=mule_rate, y=mule_pae, col sep=comma] from {csv/Nature_Flowers.csv};
	\nextgroupplot[cycle list shift=3, title style={yshift=-1.5ex,}, title={Stone Pillars}]
		\addplot[smooth, color=clr_plt1, mark=*]
			table [x=mhvs_rate, y=mhvs_pae, col sep=comma] from {csv/Urban_Stone_Pillars_Outside.csv};
		\addplot[smooth, color=clr_plt2, mark=*]
			table [x=epic_rate, y=epic_pae, col sep=comma] from {csv/Urban_Stone_Pillars_Outside.csv};
		\addplot[smooth, color=clr_plt4, mark=*]
			table [x=hevc-inter_rate, y=hevc-inter_pae, col sep=comma] from {csv/Urban_Stone_Pillars_Outside.csv};
		\addplot[smooth, color=clr_plt5, mark=*]
			table [x=mule_rate, y=mule_pae, col sep=comma] from {csv/Urban_Stone_Pillars_Outside.csv};
	\nextgroupplot[cycle list shift=4, title style={yshift=-1.5ex,}, title={Vespa}]
		\addplot[smooth, color=clr_plt1, mark=*]
			table [x=mhvs_rate, y=mhvs_pae, col sep=comma] from {csv/Mirrors_and_Transparency_Vespa.csv};
		\addplot[smooth, color=clr_plt2, mark=*]
			table [x=epic_rate, y=epic_pae, col sep=comma] from {csv/Mirrors_and_Transparency_Vespa.csv};
		\addplot[smooth, color=clr_plt4, mark=*]
			table [x=hevc-inter_rate, y=hevc-inter_pae, col sep=comma] from {csv/Mirrors_and_Transparency_Vespa.csv};
		\addplot[smooth, color=clr_plt5, mark=*]
			table [x=mule_rate, y=mule_pae, col sep=comma] from {csv/Mirrors_and_Transparency_Vespa.csv};
	\nextgroupplot[cycle list shift=5, title style={yshift=-1.7ex,}, title={Ankylosaurus}]
		\addplot[smooth, color=clr_plt1, mark=*]
			table [x=mhvs_rate, y=mhvs_pae, col sep=comma] from {csv/Studio_Ankylosaurus_Diplodocus_1.csv};
		\addplot[smooth, color=clr_plt2, mark=*]
			table [x=epic_rate, y=epic_pae, col sep=comma] from {csv/Studio_Ankylosaurus_Diplodocus_1.csv};
		\addplot[smooth, color=clr_plt4, mark=*]
			table [x=hevc-inter_rate, y=hevc-inter_pae, col sep=comma] from {csv/Studio_Ankylosaurus_Diplodocus_1.csv};
		\addplot[smooth, color=clr_plt5, mark=*]
			table [x=mule_rate, y=mule_pae, col sep=comma] from {csv/Studio_Ankylosaurus_Diplodocus_1.csv};
	\nextgroupplot[cycle list shift=6, title style={yshift=-1.5ex,}, title={Desktop}]
		\addplot[smooth, color=clr_plt1, mark=*]
			table [x=mhvs_rate, y=mhvs_pae, col sep=comma] from {csv/Studio_Desktop.csv};
		\addplot[smooth, color=clr_plt2, mark=*]
			table [x=epic_rate, y=epic_pae, col sep=comma] from {csv/Studio_Desktop.csv};
		\addplot[smooth, color=clr_plt4, mark=*]
			table [x=hevc-inter_rate, y=hevc-inter_pae, col sep=comma] from {csv/Studio_Desktop.csv};
		\addplot[smooth, color=clr_plt5, mark=*]
			table [x=mule_rate, y=mule_pae, col sep=comma] from {csv/Studio_Desktop.csv};
	\nextgroupplot[cycle list shift=7, title style={yshift=-1.5ex,}, title={Magnets}]
		\addplot[smooth, color=clr_plt1, mark=*]
			table [x=mhvs_rate, y=mhvs_pae, col sep=comma] from {csv/Studio_Magnets_1.csv};
		\addplot[smooth, color=clr_plt2, mark=*]
			table [x=epic_rate, y=epic_pae, col sep=comma] from {csv/Studio_Magnets_1.csv};
		\addplot[smooth, color=clr_plt4, mark=*]
			table [x=hevc-inter_rate, y=hevc-inter_pae, col sep=comma] from {csv/Studio_Magnets_1.csv};
		\addplot[smooth, color=clr_plt5, mark=*]
			table [x=mule_rate, y=mule_pae, col sep=comma] from {csv/Studio_Magnets_1.csv};
	\nextgroupplot[cycle list shift=8, title style={yshift=-1.5ex,}, title={Fountain}]
		\addplot[smooth, color=clr_plt1, mark=*]
			table [x=mhvs_rate, y=mhvs_pae, col sep=comma] from {csv/People_Fountain_Vincent_2.csv};
		\addplot[smooth, color=clr_plt2, mark=*]
			table [x=epic_rate, y=epic_pae, col sep=comma] from {csv/People_Fountain_Vincent_2.csv};
		\addplot[smooth, color=clr_plt4, mark=*]
			table [x=hevc-inter_rate, y=hevc-inter_pae, col sep=comma] from {csv/People_Fountain_Vincent_2.csv};
		\addplot[smooth, color=clr_plt5, mark=*]
			table [x=mule_rate, y=mule_pae, col sep=comma] from {csv/People_Fountain_Vincent_2.csv};
	\nextgroupplot[cycle list shift=9, title style={yshift=-1.7ex,}, title={Friends}]
		\addplot[smooth, color=clr_plt1, mark=*]
			table [x=mhvs_rate, y=mhvs_pae, col sep=comma] from {csv/People_Friends_1.csv};
		\addplot[smooth, color=clr_plt2, mark=*]
			table [x=epic_rate, y=epic_pae, col sep=comma] from {csv/People_Friends_1.csv};
		\addplot[smooth, color=clr_plt4, mark=*]
			table [x=hevc-inter_rate, y=hevc-inter_pae, col sep=comma] from {csv/People_Friends_1.csv};
		\addplot[smooth, color=clr_plt5, mark=*]
			table [x=mule_rate, y=mule_pae, col sep=comma] from {csv/People_Friends_1.csv};
	\nextgroupplot[cycle list shift=10, title style={yshift=-1.5ex,}, title={Color-Chart}]
		\addplot[smooth, color=clr_plt1, mark=*]
			table [x=mhvs_rate, y=mhvs_pae, col sep=comma] from {csv/ISO_and_Colour_Charts_Color_Chart_1.csv};
		\addplot[smooth, color=clr_plt2, mark=*]
			table [x=epic_rate, y=epic_pae, col sep=comma] from {csv/ISO_and_Colour_Charts_Color_Chart_1.csv};
		\addplot[smooth, color=clr_plt4, mark=*]
			table [x=hevc-inter_rate, y=hevc-inter_pae, col sep=comma] from {csv/ISO_and_Colour_Charts_Color_Chart_1.csv};
		\addplot[smooth, color=clr_plt5, mark=*]
			table [x=mule_rate, y=mule_pae, col sep=comma] from {csv/ISO_and_Colour_Charts_Color_Chart_1.csv};
	\nextgroupplot[cycle list shift=11, title style={yshift=-1.5ex,}, title={ISO-Chart}]
		\addplot[smooth, color=clr_plt1, mark=*]
			table [x=mhvs_rate, y=mhvs_pae, col sep=comma] from {csv/ISO_and_Colour_Charts_ISO_Chart_12.csv};
		\addplot[smooth, color=clr_plt2, mark=*]
			table [x=epic_rate, y=epic_pae, col sep=comma] from {csv/ISO_and_Colour_Charts_ISO_Chart_12.csv};
		\addplot[smooth, color=clr_plt4, mark=*]
			table [x=hevc-inter_rate, y=hevc-inter_pae, col sep=comma] from {csv/ISO_and_Colour_Charts_ISO_Chart_12.csv};
		\addplot[smooth, color=clr_plt5, mark=*]
			table [x=mule_rate, y=mule_pae, col sep=comma] from {csv/ISO_and_Colour_Charts_ISO_Chart_12.csv};
	\end{groupplot}
\end{tikzpicture}
\caption{\label{fig:pae_}Comparison of bitrate versus $\ell_\infty$ loss performance for tested light field coding schemes on the evaluation set. The anchors for HEVC are generated by setting QP = $\lbrace6,12,18,24\rbrace$, whereas, for MuLE $\lambda = \lbrace3,6,11,22,43,86\rbrace$ were used. For EPIC and $\ell_\infty^{4d}$DNet system, $\tau=\lbrace1..8\rbrace$ were used at the encoder to generate the data points. Comparison with $\ell_\infty$-SDNet and $\ell_\infty$-SASNet is skipped as their performance is same as $\ell_\infty^{4d}$DNet system for all the bitrates.}
\end{figure*}

We also add in the comparison group three alternative $\ell_\infty$-constrained light field compression methods to compete against the proposed $\ell_\infty^{4d}$DNet.
The first one is the recently published EPIC method for $\ell_\infty$-constrained compression of light field images without CNN-based compression error repairing
\cite{mukati2020epipolar}.  The second one modifies Zhang and Wu's $\ell_\infty$-SDNet method \cite{zhang2020ultra}, which was designed for deep soft decompression of $\ell_\infty$-coded conventional images, to repair compression errors in the SAI representation of the light field compressed by EPIC. 
The third one replicates the pipeline of the proposed coding system, but replaces $\ell_\infty^{4d}$DNet by a different light field restoration CNN, denoted as $\ell_\infty$-SASNet, introduced in \cite{jin2020learning}.  This alternative CNN 
uses a sequence of spatial-angular separable (SAS) interleaved 2D convolutions to extract features of the 4D light field.  It was originally trained for correcting inconsistencies in light field caused by independently synthesizing several views, but here retrained for repairing compression errors in the $\ell_\infty$-coded light field.  This $\ell_\infty$-SASNet method is used to contrast the $\ell_\infty^{4d}$DNet architecture with an alternative design.

%as an alternative to $\ell_\infty^{4d}$DNet

\subsection{Quantitative evaluation}
\noindent
In Fig.~\ref{fig:rd_residual}, we plot the rate-distortion curves of the five competing methods. As clearly demonstrated, our $\ell_\infty^{4d}$DNet method achieves the highest PSNR in the comparison group for each of the 12 light field test images.
Among the $\ell_\infty$-constraint compression systems, the $\ell_\infty$-SDNet method is the poorest performer as it only uses the correlations in the axial direction of light field in the restoration process.
The other two deep decompression networks,
$\ell_\infty$-SASNet and $\ell_\infty^{4d}$DNet, both extract 4D light field features without explicitly using 4D convolution kernels.  However, the proposed network $\ell_\infty^{4d}$DNet consistently outperforms $\ell_\infty$-SASNet because it explores the 4D correlations of light field images more thoroughly.

We also present, in Table \ref{tab:bjontegaard}, the results in Bjøntegaard delta peak signal-to-noise ratio (BD-PSNR). The table shows that $\ell_\infty^{4d}$DNet beats
HEVC by $3.21$ dBs in BD-PSNR, whereas most of the other methods perform worse than HEVC.

Finally, we compare the $\ell_\infty$ performances of the four competing methods in Fig.~\ref{fig:pae_}.  Note that $\ell_\infty$-SDNet and $\ell_\infty$-SASNet have the same $\ell_\infty$ error as the proposed $\ell_\infty^{4d}$DNet.

By reading Fig.~\ref{fig:rd_residual} and Fig.~\ref{fig:pae_} in contrast, one can see that $\ell_\infty^{4d}$DNet performs the best in $\ell_2$ error metric, and the second best in $\ell_\infty$. As we strived for in the design of $\ell_\infty^{4d}$DNet, the new method obtains a good balance between the average and minmax error criteria.  For example, $\ell_\infty^{4d}$DNet improves
the PSNR of $\ell_\infty$-constrained EPIC by $5.61$ dBs on average; this gain in the $\ell_2$ metric over the $\ell_\infty$-constrained EPIC is achieved at the cost of relaxing the $\ell_\infty$ bound from $\tau$ to $2\tau$. Here it should be stressed that $\ell_\infty^{4d}$DNet, although with $\ell_\infty$-constraint, beats the two lossy codecs HEVC and MuLE even in $\ell_2$ metric when the bit rate is above 0.1 bpp on average or the minimum acceptable quality is above 40 dBs. And at the same time, as shown in Fig.~\ref{fig:pae_},  $\ell_\infty^{4d}$DNet enjoys a large $\ell_\infty$ advantage over HEVC and MuLE.

\begin{figure*}
  \centering
  \begin{subfigure}[b]{0.2\textwidth}
  \caption*{MuLE (0.135 bpp)}
  \vspace{-0.5em}
  \includegraphics[width=\textwidth]{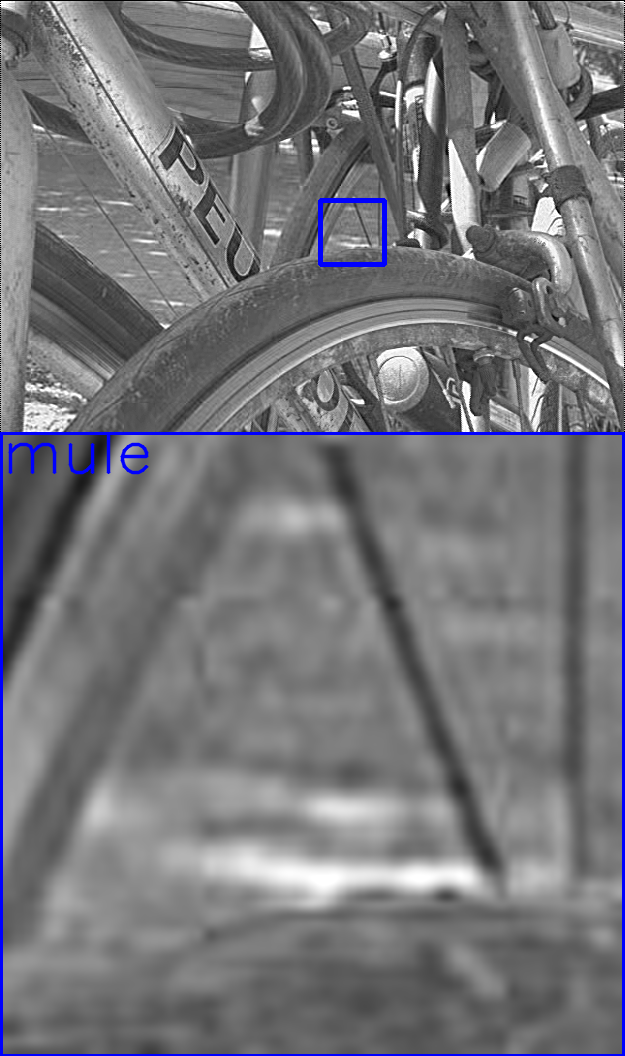}
  \caption*{$35.02dB/\norm{e}_\infty = 68$}%
  \end{subfigure}%
  \begin{subfigure}[b]{0.2\textwidth}
  \caption*{HEVC (0.126 bpp)}
  \vspace{-0.5em}
  \includegraphics[width=\textwidth]{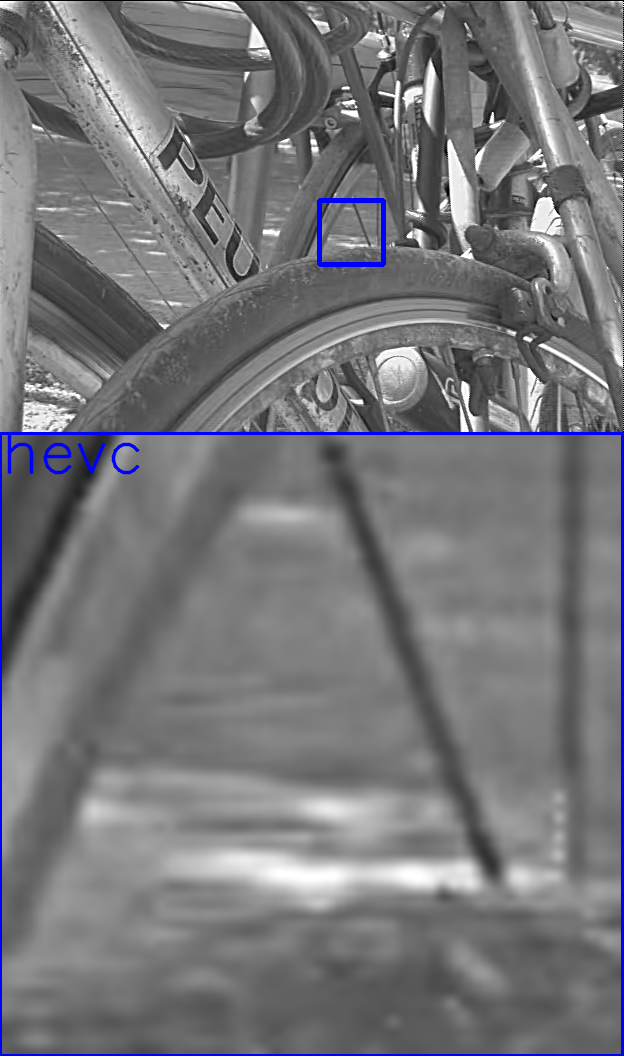}
  \caption*{$38.14dB/\norm{e}_\infty = 23$}%
  \end{subfigure}%
  \begin{subfigure}[b]{0.2\textwidth}
  \caption*{EPIC (0.128 bpp)}
  \vspace{-0.5em}
  \includegraphics[width=\textwidth]{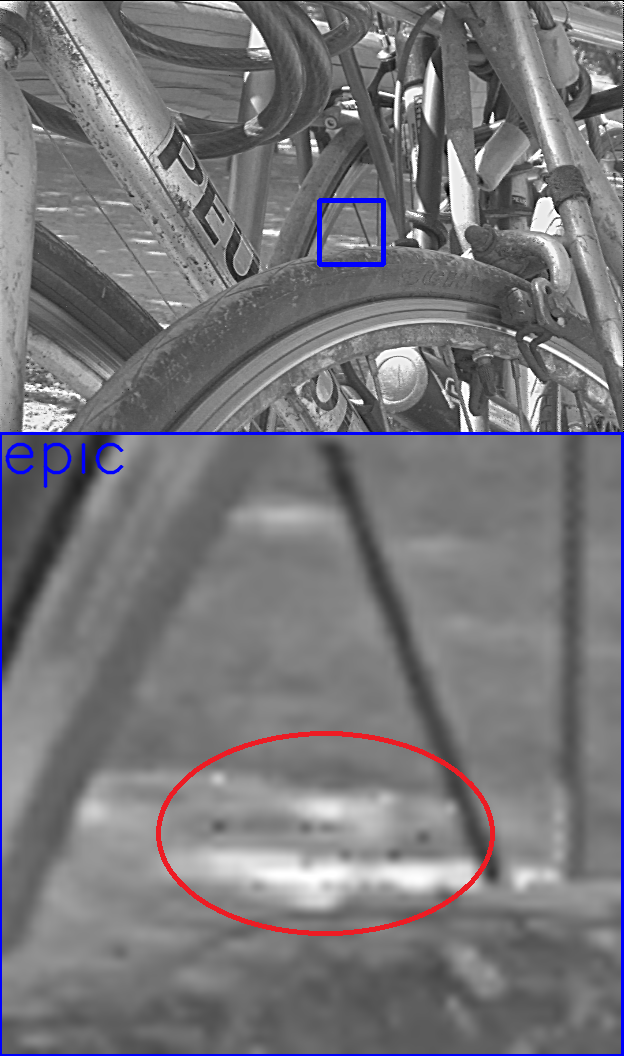}
  \caption*{$40.23dB/\norm{e}_\infty = 10$}%
  \end{subfigure}%
  \begin{subfigure}[b]{0.2\textwidth}
  \caption*{$\ell_\infty^{4d}$DNet (0.128 bpp)}
  \vspace{-0.5em}
  \includegraphics[width=\textwidth]{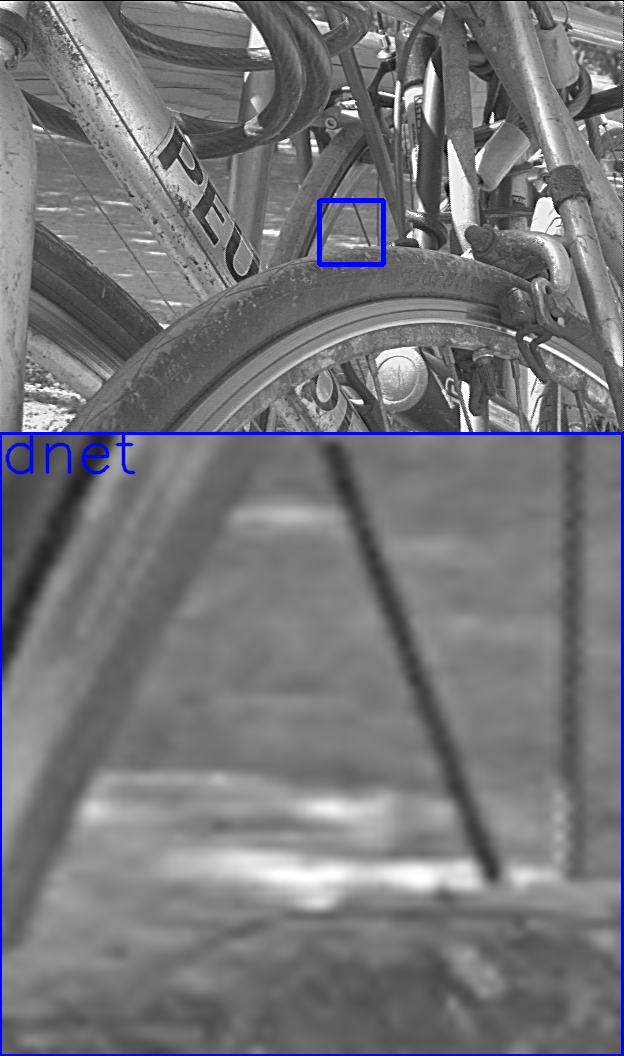}
  \caption*{$44.69dB/\norm{e}_\infty = 16$}%
  \end{subfigure}%
  \begin{subfigure}[b]{0.2\textwidth}
  \caption*{Ground Truth}
  \vspace{-0.5em}
  \includegraphics[width=\textwidth]{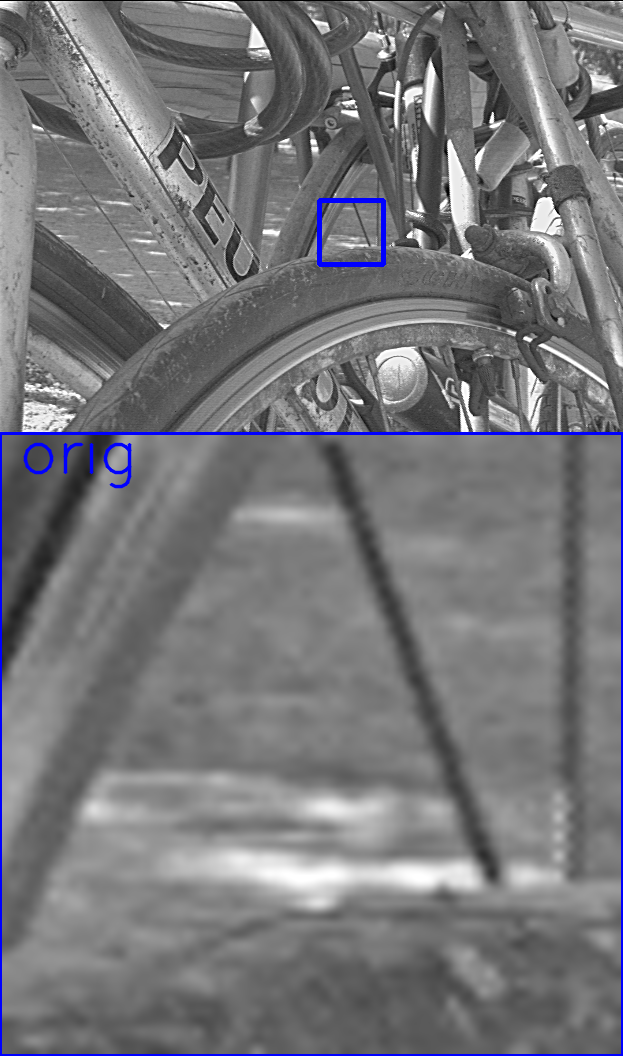}
  \caption*{PSNR/$\norm{e}_\infty$}%
  \end{subfigure}%
  \vspace{-1em}
  \caption{\label{fig:visual_bikes} Visual comparisons of competing methods.
  In the close-up views, note the severely blurred tire thread by MuLE and HEVC in contrast to $\ell_\infty^{4d}$DNet, and also the speckle compression artifacts (in red ellipse) of EPIC that are removed by $\ell_\infty^{4d}$DNet.  The images are the top-central light field views reconstructed by the tested methods.}
\end{figure*}

\begin{figure*}
  \centering
  \begin{subfigure}[b]{0.2\textwidth}
  \caption*{MuLE (0.146 bpp)}
  \vspace{-0.5em}
  \includegraphics[width=\textwidth]{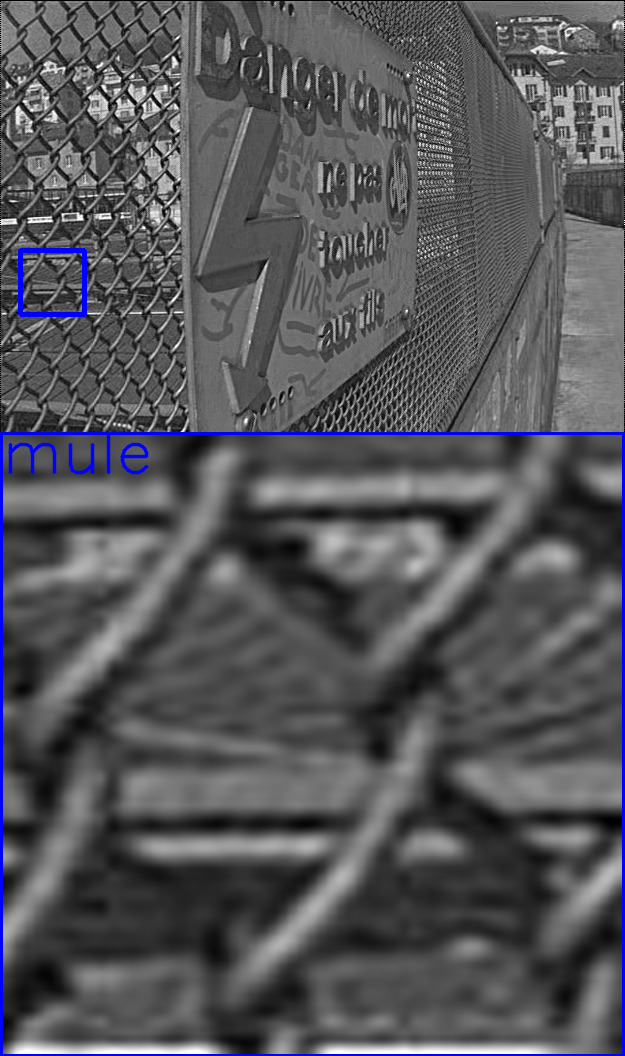}
  \caption*{$33.98dB/\norm{e}_\infty = 53$}%
  \end{subfigure}%
  \begin{subfigure}[b]{0.2\textwidth}
  \caption*{HEVC (0.146 bpp)}
  \vspace{-0.5em}
  \includegraphics[width=\textwidth]{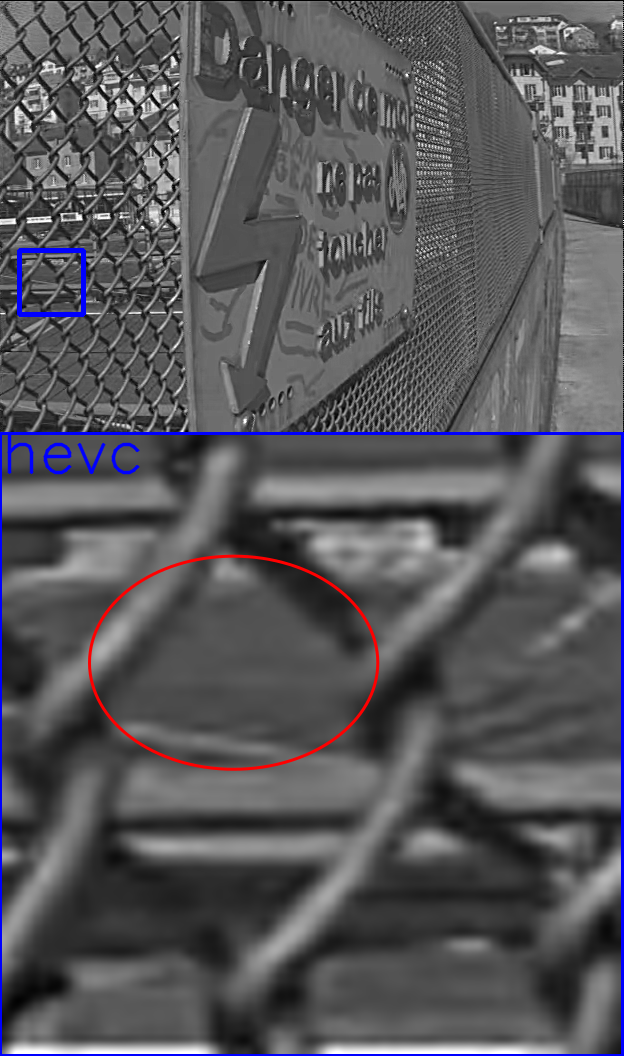}
  \caption*{$36.19dB/\norm{e}_\infty = 29$}%
  \end{subfigure}%
  \begin{subfigure}[b]{0.2\textwidth}
  \caption*{EPIC (0.156 bpp)}
  \vspace{-0.5em}
  \includegraphics[width=\textwidth]{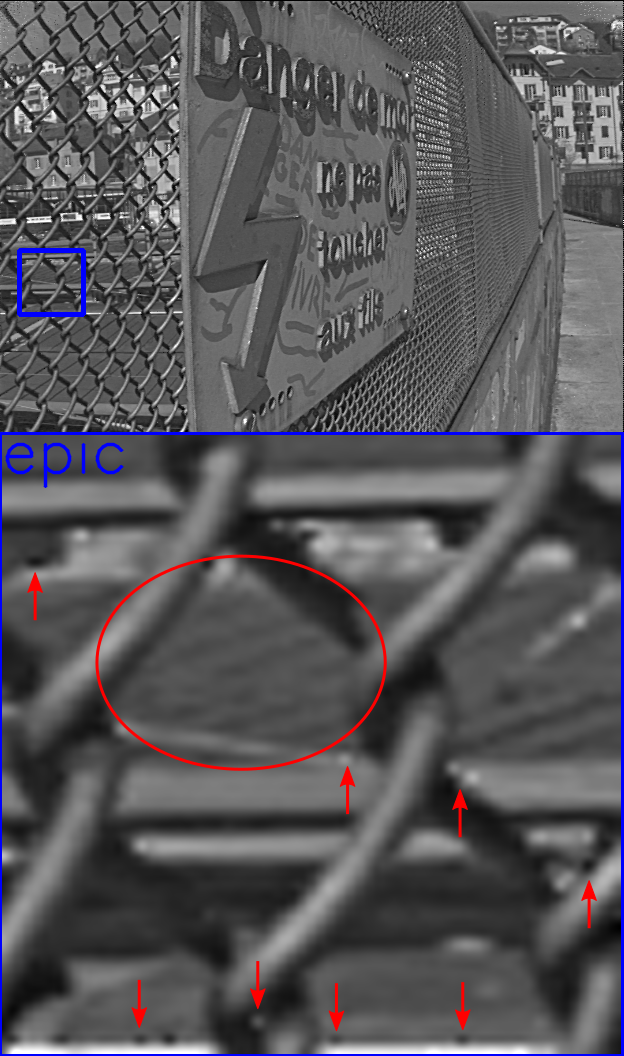}
  \caption*{$40.23dB/\norm{e}_\infty = 10$}%
  \end{subfigure}%
  \begin{subfigure}[b]{0.2\textwidth}
  \caption*{$\ell_\infty^{4d}$DNet (0.156 bpp)}
  \vspace{-0.5em}
  \includegraphics[width=\textwidth]{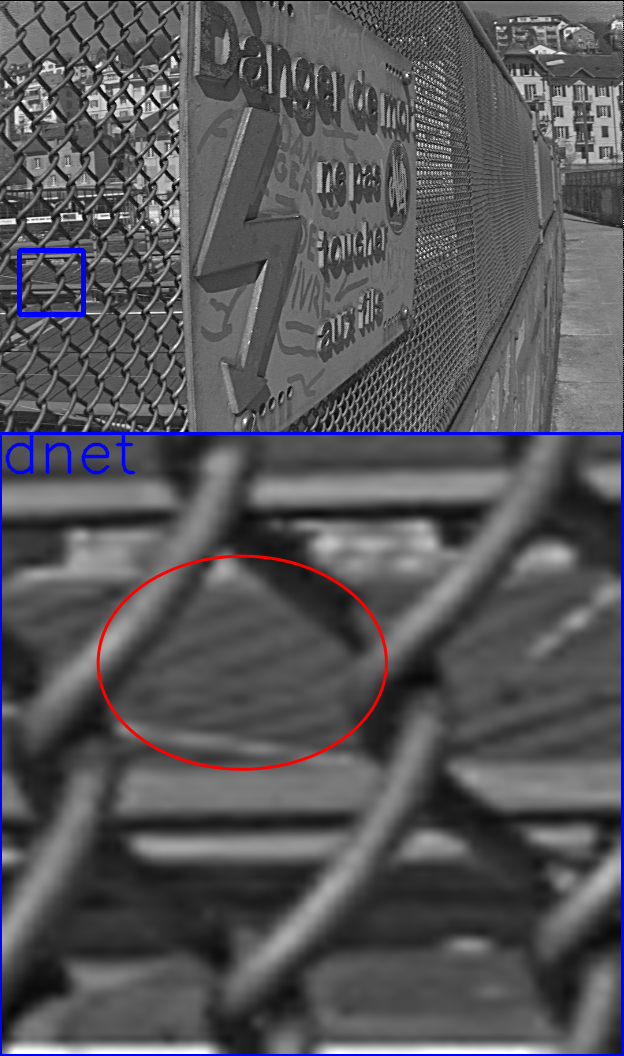}
  \caption*{$44.57dB/\norm{e}_\infty = 13$}%
  \end{subfigure}%
  \begin{subfigure}[b]{0.2\textwidth}
  \caption*{Ground Truth}
  \vspace{-0.5em}
  \includegraphics[width=\textwidth]{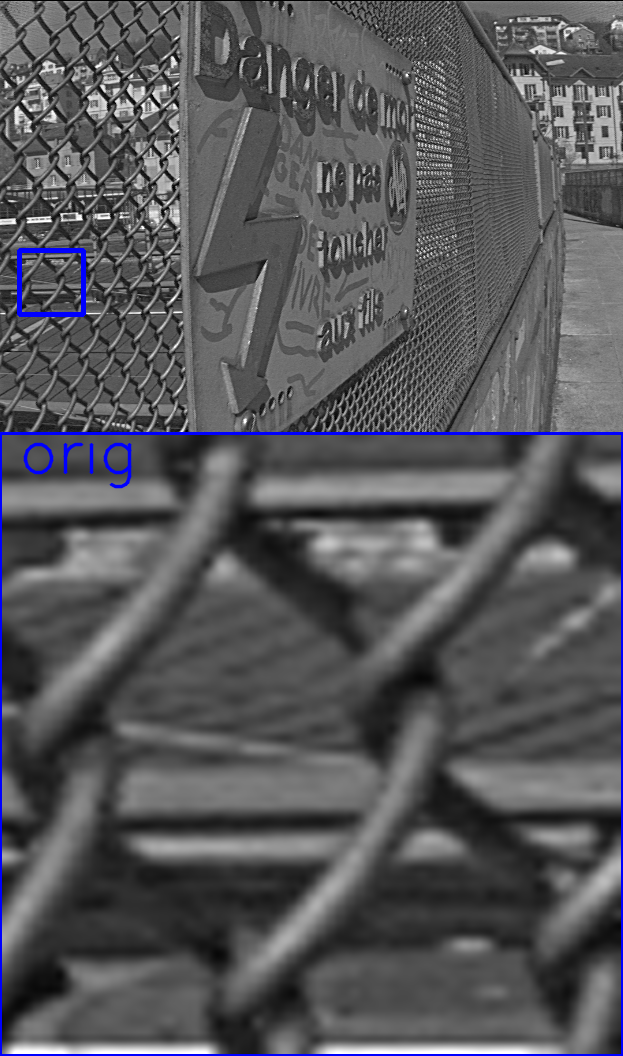}
  \caption*{PSNR/$\norm{e}_\infty$}%
  \end{subfigure}%
  \vspace{-1em}
  \caption{\label{fig:visual_danger} Visual comparisons of different methods. Note the flattening of the roof pattern (red ellipse region) in the HEVC result and the speckle compression artifacts of EPIC (marked by red arrows), which are nonexistent in the output image of $\ell_\infty^{4d}$DNet. The images are the top-central light field views reconstructed by the tested methods.}
\end{figure*}

% \subsection{Visual comparison}
\subsection{Qualitative evaluation}
\noindent
In Figs.~\ref{fig:visual_bikes} and \ref{fig:visual_danger}, we compare different methods in visual quality for the approximately same bit rates.  
%To demonstrate the superiority of the $\ell_\infty^{4d}$DNet over other methods, unsharp masking technique is applied to enhance all the images shown in the two figures.
The high frequency details in Fig.~\ref{fig:visual_bikes}, such as the tire threads and the wheel spokes, are smoothed out by MuLE and HEVC.  Although EPIC preserves the high frequency details better, it suffers from impulse compression artifacts (see the red ellipse region) in the close-up view.
Similarly, in Fig.~\ref{fig:visual_danger}, HEVC smooths out high frequency details, while EPIC is susceptible to speckle compression artifacts.  MuLE is the worst performer among the four competing methods.  Only the proposed $\ell_\infty^{4d}$DNet method achieves perceptually lossless reconstruction in both cases.
																											
% Again, only the $\ell_\infty^{4d}$DNet is flawless.

\subsection{Comparison of complexities}
\noindent
Recall that one of the motivations of this research is to devise a low-complexity encoder for light field images.
Now let us compare the encoder complexity of $\ell_\infty^{4d}$DNet against with those of HEVC and MuLE.
For instance, to encode a light field, depending on the quality level, HEVC requires 30 to 60 minutes on HP ZBook 15 with 16 GB RAM; MuLE requires 15 to 30 minutes. In contrast,
the $\ell_\infty$-based DPCM encoder, by shifting computation burdens to $\ell_\infty^{4d}$DNet, only requires 30 to 60 seconds to encode a light field.

Also, we state the complexities of the three deep decompression networks, $\ell_\infty^{4d}$DNet, $\ell_\infty$-SASNet and $\ell_\infty$-SDNet, which require $1.02$M, $1.24$M and $2.17$M network parameters, respectively. 

\begin{figure*}[!t]
     \centering
     \begin{subfigure}[b]{0.18\textwidth}
         \centering
         \includegraphics[width=\textwidth]{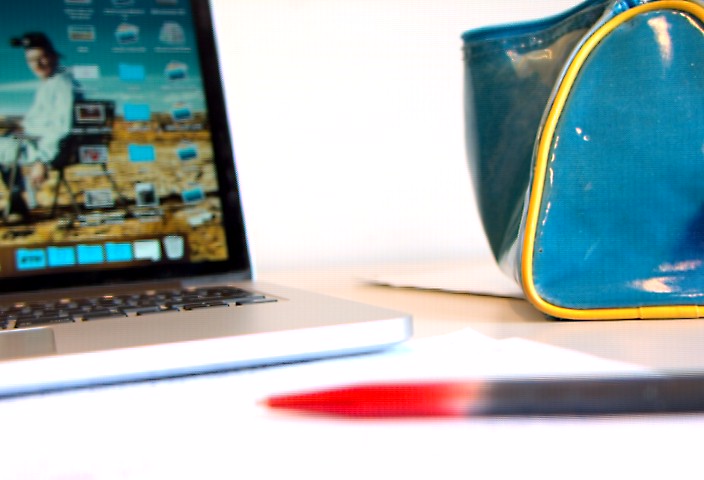}
         \\
         \footnotesize{(a) A light field scene}
         \label{fig:repr_weight:thumbnail}
     \end{subfigure}
     \hfill
     \begin{subfigure}[b]{0.18\textwidth}
         \centering
         \includegraphics[width=\textwidth]{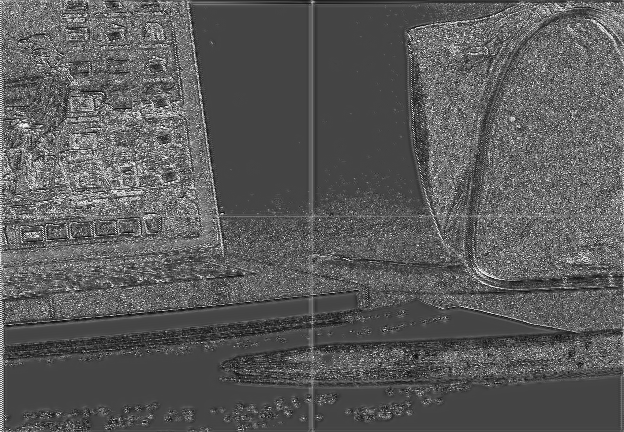}
         \\
         \footnotesize{(b) MI}
         \label{fig:repr_weight:MI}
     \end{subfigure}
     \hfill
     \begin{subfigure}[b]{0.18\textwidth}
         \centering
         \includegraphics[width=\textwidth]{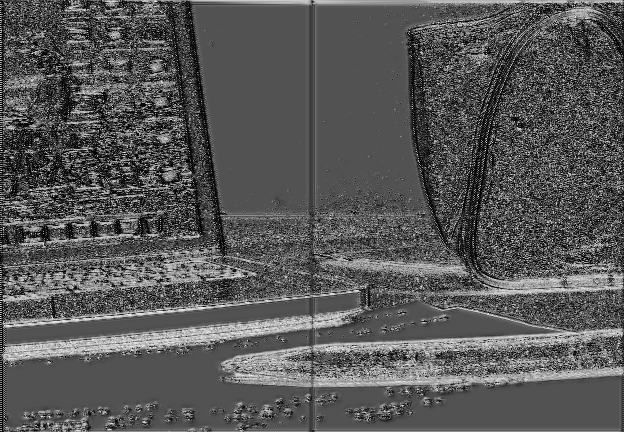}
         \\
         \footnotesize{(c) EPI-H}
         \label{fig:repr_weight:EPIH}
     \end{subfigure}
     \hfill
     \begin{subfigure}[b]{0.18\textwidth}
         \centering
         \includegraphics[width=\textwidth]{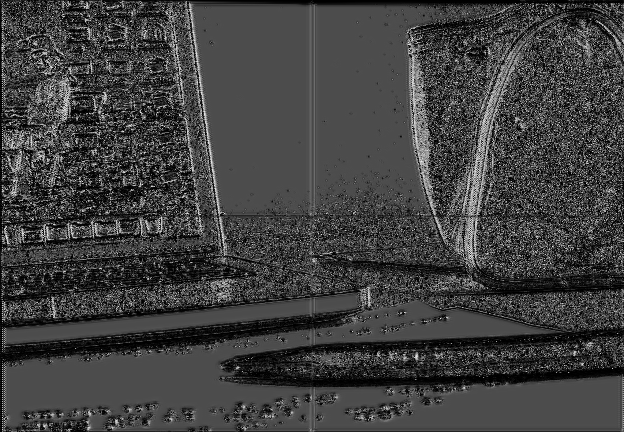}
         \\
         \footnotesize{(d) EPI-V}
         \label{fig:repr_weight:EPIV}
     \end{subfigure}
     \hfill
     \begin{subfigure}[b]{0.18\textwidth}
         \centering
         \includegraphics[width=\textwidth]{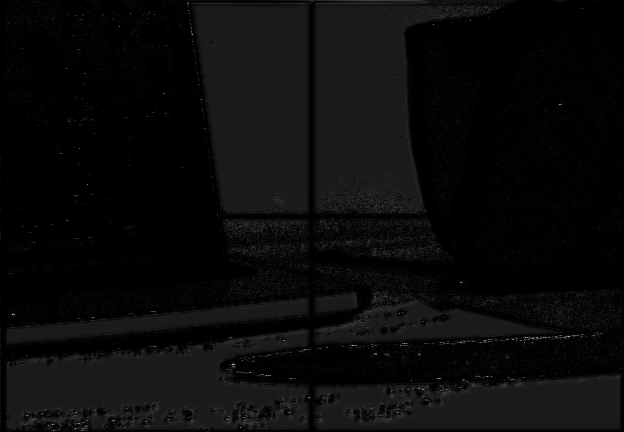}
         \\
         \footnotesize{(e) SAI}
         \label{fig:repr_weight:SAI}
     \end{subfigure}
    \caption{Weight maps $W_k$ of the four 2D representations calculated by GateNet, for the central view of the light field ``Desktop'' from the evaluation set.}
    \label{fig:repr_weight}
\end{figure*}

\section{Ablation Studies}
\label{sec:ablation}
\noindent
In this section, we conduct two ablation experiments to analyze and justify the proposed design of $\ell_\infty^{4d}$DNet.

\begin{figure*}[!ht]
\centering
\includegraphics[width=0.8\textwidth]{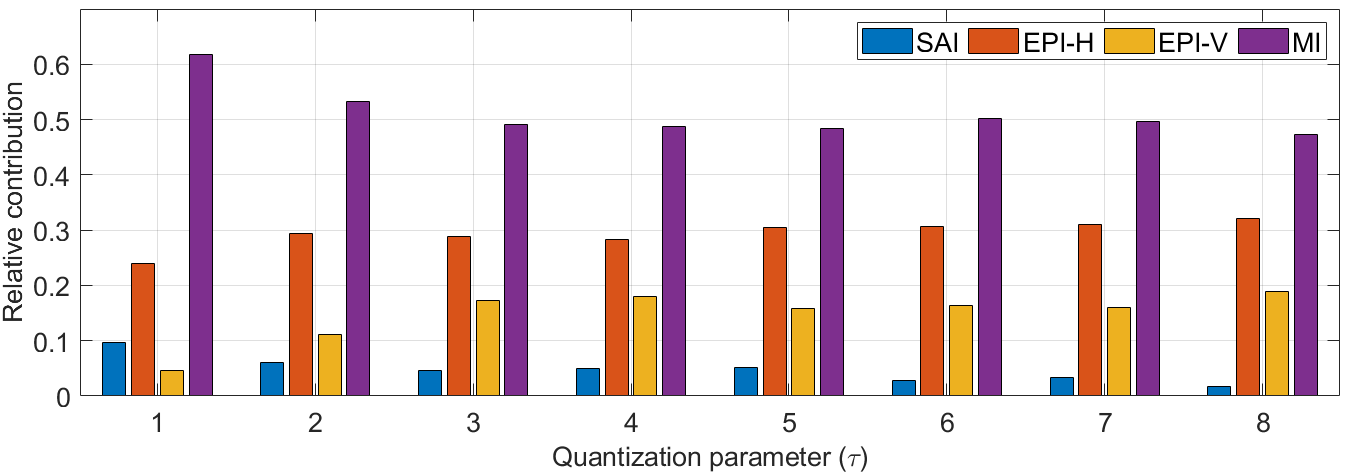}
\caption{\label{fig:bargraph_poe} Relative importance of different 2D representations in the restoration of compressed light field.}
\vspace{-1em}
\end{figure*}

\subsection{Roles of different 2D representations}
\noindent
To better understand the role each 2D representation plays in light field reconstruction, we examine the spatial weight maps calculated by the GateNet.  Fig.~\ref{fig:repr_weight} shows, in an example, a light field image and the spatial weight maps of the four 2D representations, SAI, EPI-H, EPI-V, and MI.
As one can see, spatial regions with zero disparity, result in flat MI representation, for e.g. laptop screen, bag, etc; hence the GateNet locally assigns a larger weight to MI estimates as shown in Fig.~\ref{fig:repr_weight}b.
A clear contrast can be observed between EPI-H and EPI-V weights in Fig.~\ref{fig:repr_weight}c and Fig.~\ref{fig:repr_weight}d for horizontal and vertical edges of the scene, respectively.
A large contribution is made by EPI-H estimates for horizontal edges, i.e., lower prediction errors of EPI-H in the presence of horizontal occlusion. Likewise, the contribution of EPI-V is more significant for vertical edges.
Due to their low prediction power, the SAI estimates are given the least significance as illustrated in Fig.~\ref{fig:repr_weight}e.
The above empirical observations support the arguments made in Section \ref{sec:gatenet}.

In Fig.~\ref{fig:bargraph_poe}, we plot the relative contributions of the four 2D representations in the restoration of the compressed light field images, in relation to the $\ell_\infty$ distortion bound $\tau$.
The relative contribution of a 2D representation $k$, $k \in \lbrace \text{SAI}, \text{EPI-H}, \text{EPI-V}, \text{MI}\rbrace$, is calculated as follows:
\begin{equation*}
C_k = \dfrac{\sum_x W_k(x)}{\sum_k \sum_x W_k(x)},
\end{equation*}
\noindent
where $W$ is the spatial weight map, $x$ indexes the pixel locations.
The histogram plots in Fig.~\ref{fig:bargraph_poe} are averaged over the evaluation set.  The rank of the relative contribution or importance of the four 2D representations remains the same for $\tau \geq 2$.  For the case of $\tau = 1$, the SAI representation has relatively higher importance than for larger $\tau$ values.

\subsection{Impact of dropping 2D representations}
\noindent
In the proposed design of $\ell_\infty^{4d}$DNet, we use four 2D representations of the 4D light field, SAI, EPI-H, EPI-V, and MI. A tantalizing question is what if only a subset of these four 2D representations is used for the sake of complexity reduction.
We design and train two $\ell_\infty^{4d}$DNet variants, each utilizing a different subset of the four 2D representations, and compare them with the proposed $\ell_\infty^{4d}$DNet.
The first variant $\ell_\infty^{4d}$DNet-$1R$ utilizes only one 2D representation, which is a highly predictive MI representation, hoping to retain reconstruction quality while drastically reducing the model size.  On the other hand, the second variant $\ell_\infty^{4d}$DNet-$3R$ works on three representations MI, EPI-H, EPI-V, dropping the less predictable and yet resource-demanding SAI representation.

\begin{table}
\centering
\caption{\label{tab:repr_comp} Average performance gains in PSNR (dB), over the quality of the decompressed light fields in the evaluation set, of the proposed
$\ell_\infty^{4d}$DNet variants using different numbers of 2D representations for $1 \leq \tau \leq 8$.}
\begin{tabularx}{0.37\textwidth}{r|SSSS}
\hline
 \multicolumn{1}{r|}{$\tau$} & \multicolumn{1}{c}{$\ell_\infty^{4d}$DNet-$1R$} &  \multicolumn{1}{c}{$\ell_\infty^{4d}$DNet-$3R$} & \multicolumn{1}{c}{$\ell_\infty^{4d}$DNet(-$4R$)}\\
% \multicolumn{1}{c}{$\ell_\infty$-SAINet} &
\hline
\hline
1 & 1.50 & 2.02 & 2.38\\ %0.77 &
2 & 2.95 & 3.84 & 4.10\\ %& 1.44
3 & 3.67 & 4.68 & 5.01\\%& 1.70
4 & 4.19 & 5.42 & 5.66\\%& 1.90
5 & 4.70 & 6.05 & 6.25\\%& 2.15
6 & 5.27 & 6.74 & 6.87\\%& 2.47
7 & 5.42 & 6.94 & 7.03\\%& 2.41
8 & 5.85 & 7.40 & 7.58\\%& 2.70
\hline
Avg. & 4.19 & 5.39 & 5.61\\%& 1.94
\hline
\end{tabularx}
\vspace{-2em}
\end{table}%

The comparison results in PSNR gains are reported in Table~\ref{tab:repr_comp}.
Unsurprisingly, the $\ell_\infty^{4d}$DNet, which uses four light field representations for light field decompression, beats the other two variants of reduced complexity.
Although the first variant $\ell_\infty^{4d}$DNet-$1R$ offers significant reduction in computational complexity, it lags behind the $\ell_\infty^{4d}$DNet in PSNR by $1.42$dB on average.
However, the second variant $\ell_\infty^{4d}$DNet-$3R$ performs very close to the proposed $\ell_\infty^{4d}$DNet in PSNR, losing $0.22$dB on average.

To quantify computational complexities of the proposed $\ell_\infty^{4d}$DNet and the two variants, we report model size and inference time in Table~\ref{tab:model_comp}.
The first variant $\ell_\infty^{4d}$DNet-$1R$ roughly reduces model size by $9\times$ and inference time by $5\times$, however it considerably degrades the restoration performance.
On the other hand, the second variant $\ell_\infty^{4d}$DNet-$3R$ tries to minimize the performance gap by exploiting two more representations, EPI-H and EPI-V, at a cost of $5\times$ more modeling parameters than the first variant $\ell_\infty^{4d}$DNet-$1R$.

\begin{table}[ht]
\centering
\caption{\label{tab:model_comp} Comparison of proposed network variants, alternate network designs, and the proposed network, using model size and inference times }
\begin{tabularx}{0.32\textwidth}{r|SSSS}
\hline
 \multicolumn{1}{c|}{} & \multicolumn{1}{c}{Model size}  & \multicolumn{1}{c}{Inference time}\\
%FLIF
\hline
\hline
$\ell_\infty^{4d}$DNet-$1R$ & 0.25M & 0.02s\\
$\ell_\infty^{4d}$DNet-$3R$ & 1.13M & 0.06s\\
% BlendNet & 1.24M & 0.02s\\
% MedianNet & 1.81M & 0.09s\\
$\ell_\infty^{4d}$DNet(-$4R$) & 2.17M & 0.10s\\
\hline
\end{tabularx}
\vspace{-1.5em}
\end{table}%

\section{Conclusion}
\label{sec:conclusion}
\noindent
In this paper, we proposed a highly efficient
%$\ell_\infty^{4d}$DNet based
$\ell_\infty$-constrained light field compression system, which uses the proposed soft decompression network $\ell_\infty^{4d}$DNet to remove the near-lossless compression artifacts introduced by the EPIC codec, by capitalizing on the 4D structure of light fields. The $\ell_\infty^{4d}$DNet reconstructs a higher quality light field by fusing multiple estimates of the compression error. These estimates are made by the four light field modeling experts, each of which uses a different 2D representation of the 4D light field.  The fusion weights are determined by a gating network.
The proposed $\ell_\infty^{4d}$DNet based compression system is shown to achieve superior rate-distortion performance to the state-of-the-art lossy compression schemes, while also offering a tight error bound for every pixel.
% \section*{Acknowledgment}

% \nocite{*}
\bibliographystyle{IEEEtran}
\bibliography{references}

\begin{IEEEbiography}[{\includegraphics[width=1in,height=1.25in,clip,keepaspectratio]{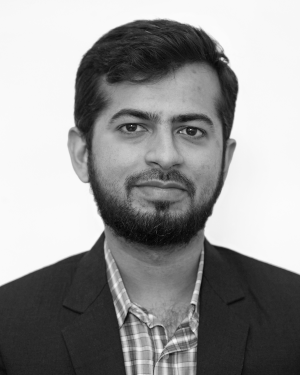}}]{M. Umair Mukati} received the B.E. degree in Industrial Electronics from NED University of Engineering \& Technology, Pakistan
and the M.S. degree in Electronics with specialization in computational imaging from Istanbul Medipol University, Turkey in 2014 and 2017, respectively.
During his masters he worked on improving the spatial resolution and virtually extending the aperture size of light fields.
He is currently pursuing the Ph.D. degree from Department of Photonics Engineering in Technical University of Denmark under Coding and Visual Communication group.
He is a part of this group since 2018.
As an Early Stage Researcher in the EU MSCA ITN RealVision, his goal is to encode light field data for flexible access and processing.
His research interests includes exploring unconventional methods for light field coding and compression, novel view synthesis and enhancing light field capabilities.
\end{IEEEbiography}

\begin{IEEEbiography}[{\includegraphics[width=1in,height=1.25in,clip,keepaspectratio]{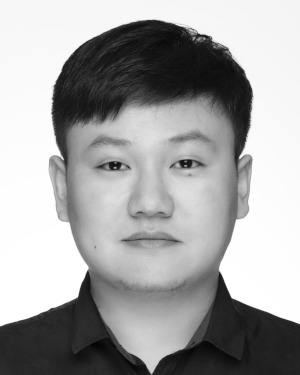}}]{Xi Zhang}
received the B.Sc. degree in mathematics and physics basic science from University of Electronic Science and Technology of China (UESTC) in 2015. He is currently pursuing the Ph.D. degree from the Department of Electronic Engineering at Shanghai Jiao Tong University (SJTU). In 2019, He was also a visiting Ph.D. student with the Department of Electrical and Computer Engineering at McMaster University, Hamilton, ON, Canada. His research interests include image/video compression, low-level vision and cognitive computing.
\end{IEEEbiography}

\begin{IEEEbiography}[{\includegraphics[width=1in,height=1.25in,clip,keepaspectratio]{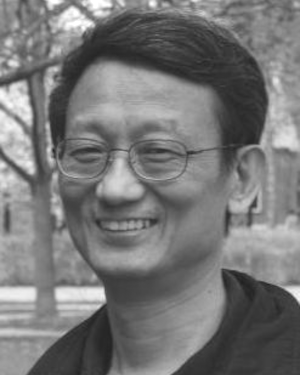}}]{Xiaolin Wu}
(Fellow, IEEE) received his B.Sc. from Wuhan University, China, in 1982, and Ph.D. from the University of Calgary, Canada, in 1988, both in computer science. He started his academic career in 1988, and was a faculty member at Western University, Canada, New York Polytechnic University (NYU-Poly), USA, and is currently with McMaster University, Canada, where he is a distinguished engineering professor and holds an NSERC Senior Industrial Research Chair. His research interests include image processing, data compression, digital multimedia, low-level vision and network-aware visual communication. He has authored and coauthored more than 300 research papers and holds four patents in these fields. He is an IEEE Fellow, an associated editor of IEEE Transactions on Image Processing, and past associated editor for the IEEE Transactions on Multimedia. He served on technical committees of many IEEE international conferences/workshops on image processing, multimedia, data compression, and information theory.
\end{IEEEbiography}

\begin{IEEEbiography}[{\includegraphics[width=1in,height=1.25in,clip,keepaspectratio]{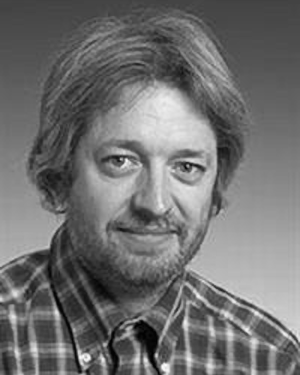}}]{Søren Forchhammer} (Member, IEEE) received the M.S. degree in engineering and the Ph.D. degree from the Technical University of Denmark, Kongens Lyngby, Denmark, in 1984 and 1988, respectively. He is currently a Professor with DTU Fotonik, Technical University of Denmark, where he has been since 1988. He is also the head of the Coding and Visual Communication Group with DTU Fotonik. He is currently coordinator of the EU MSCA ITN RealVision. His interests include source coding, image and video coding, distributed source coding, processing for image displays, 2D information theory, coding for optical communication and visual communications.
\end{IEEEbiography}

\end{document}